\documentclass[prd,amssymb,amsmath,showpacs,nofootinbib,superscriptaddress]{revtex4}
\usepackage{graphicx}
\allowdisplaybreaks[4]

\usepackage{latexsym}

\begin{document}

\unitlength=1mm

% Greek Letters
\def\a{{\alpha}}
\def\b{{\beta}}
\def\d{{\delta}}
\def\D{{\Delta}}
\def\e{{\epsilon}}
\def\g{{\gamma}}
\def\G{{\Gamma}}
\def\k{{\kappa}}
\def\l{{\lambda}}
\def\L{{\Lambda}}
\def\m{{\mu}}
\def\n{{\nu}}
\def\w{{\omega}}
\def\O{{\Omega}}
\def\S{{\Sigma}}
\def\s{{\sigma}}
\def\t{{\tau}}
\def\th{{\theta}}
\def\x{{\xi}}

\def\ol#1{{\overline{#1}}}

%slash's
\def\Dslash{D\hskip-0.65em /}
\def\dslash{{\partial\hskip-0.5em /}}
\def\vslash{{\rlap \slash v}}
\def\qbar{{\overline q}}

% Jargon
\def\CPT{{$\chi$PT}}
\def\QCPT{{Q$\chi$PT}}
\def\PQCPT{{PQ$\chi$PT}}
\def\tr{\text{tr}}
\def\str{\text{str}}
\def\diag{\text{diag}}
\def\order{{\mathcal O}}
\def\vit{{\it v}}
\def\vD{\vit\cdot D}
\def\am{\alpha_M}
\def\bm{\beta_M}
\def\gm{\gamma_M}
\def\smb{\sigma_M}
\def\smt{\overline{\sigma}_M}
\def\tb{{\tilde b}}

\def\c#1{{\mathcal #1}}

% Fields
\def\Bbar{\overline{B}}
\def\Tbar{\overline{T}}
\def\cBbar{\overline{\cal B}}
\def\cTbar{\overline{\cal T}}
\def\pq{(PQ)}

\def\eqref#1{{(\ref{#1})}}

%\preprint{DUKE-TH-04-XXX}
\preprint{NT@UW 04-023}

\title{Strong Isospin Breaking of the Nucleon and Delta Masses on the Lattice}
\author{Brian C.~Tiburzi}
\email[]{bctiburz@phy.duke.edu}
\affiliation{Department of Physics Box 90305,
Duke University,
Durham, NC 27708-0305}

\author{ Andr\'e Walker-Loud} 
\email[]{walkloud@u.washington.edu}
\affiliation{Department of Physics Box 351560,
University of Washington,
Seattle, WA 98195-1560}

\date{\today}

\begin{abstract}
Strong isospin breaking in the spectrum of the nucleons and deltas can
be studied in lattice QCD with the help of chiral perturbation theory.
At leading order in the chiral expansion, the mass splittings
between the proton and neutron and between the deltas are linear in the
quark mass difference.  The next-to-leading order contributions to
these splittings vanish even away from the strong-isospin limit.
Therefore, any non-linear quark mass dependence of these mass splittings is
a signal of the next-to-next-to-leading order mass
contributions, thus providing access to LECs at this order.
We determine the mass splittings of the
nucleons and deltas in two-flavor, heavy baryon chiral perturbation
theory to next-to-next-to-leading order.  
We also derive expressions for the nucleon and delta masses in
partially quenched chiral perturbation theory to the same order.
The resulting mass expressions will be useful both for the
extrapolation of lattice data on baryon masses, and for the study of
strong isospin breaking.
\end{abstract}

\maketitle

%
%
%
%
%
%
%	Introduction
%
%
%
%
%
%

\section{Introduction}
Understanding nuclear physics from first principles is
a long standing challenge.  Quantum chromodynamics (QCD)
is the well established theory of strong interactions that
describes the underlying interactions of colored quarks and gluons which bind
together to form the color neutral hadrons observed in nature:
nucleons, pions, \emph {etc}.  The equations of motion for the quarks and
gluons are non-linear, and at energy scales relevant for nuclear
physics, $E \sim 1$~GeV, the theory is non-perturbative.  Thus the description of
hadrons from first principles remains elusive.  One can
write down an effective theory of QCD, chiral perturbation
theory (\CPT)~\cite{Gasser:1983yg}, which provides a model-independent
description of the low energy interactions of the hadrons by
preserving the symmetries of QCD.  \CPT\ is written in terms of the
pions (pseudo-Goldstone bosons which appear from spontaneous chiral
symmetry breaking) and their interactions with baryonic states, 
like the nucleons and delta-resonances.
As the pions are the lightest particles in the theory, they 
dominate the low-energy interactions of the hadrons.
In \CPT, short distance physics is encoded in an infinite tower of
higher dimensional operators, whose coefficients are low-energy
constants (LECs).  The utility of \CPT\ is that these higher
dimensional operators are suppressed by appropriate powers of the
cutoff, where operators with higher inverse powers of the cutoff are
higher order in 
the expansion.  Thus, to a given order, one needs to keep only a
finite number of operators, 
and the theoretical error associated with a given calculation is
indicated by the size of omitted terms.
Symmetry constrains the number of operators which
enter at a given order, but not the numerical values of their
coefficients.  Thus the LECs must be fit by comparing either
to experiment or lattice QCD.

Lattice QCD is a tool one can use to
numerically calculate QCD observables from first principles---the mass
spectrum of baryons and mesons being notable examples.  Due
to computational restrictions, however, the light quarks must presently be
simulated with unphysically large masses.  So that observables
calculated on the lattice can be extrapolated from the larger quark
masses used in lattice simulations to the the physical quark 
masses,  one requires an understanding of the quark mass dependence of
QCD.  For low-energy QCD, \CPT\ is precisely what is
needed, as the quark masses explicitly break chiral
symmetry.  For the up and down quarks, this breaking can be treated
perturbatively; thus, these quark masses enter \CPT\ as parameters of
the theory, allowing QCD observables to be calculated as a systematic
expansion in powers of the quark masses.  Consequently lattice QCD can be used
in conjunction with \CPT\ to first determine the LECs relevant to a
given observable, and once these LECs are determined, extrapolate to
make rigorous first principles predictions of QCD observables.

Lattice QCD calculations of the baryon masses began in the early
1980's but until recently, unquenched lattice calculations have been
limited to quark masses such that the pion has been rather heavy,
$m_\pi \gtrsim 500$~MeV~\cite{Bernard:2001av,AliKhan:2001tx,
  Aoki:2002uc,Zanotti:2003fx,Bernard:2003rp,Aubin:2004wf,
  Toussaint:2004cj, Renner:2004ck, Heller:2004bm}.
There are a handful of fits to the lattice data on baryon masses using
\CPT~\cite{Leinweber:2003dg,Procura:2003ig,AliKhan:2003cu,Meissner:2005mb}.
However, the viability of \CPT\ with such large pion masses is
questionable~%
\cite{Bernard:2002yk,Beane:2004ks}.  Ideally, one would like to see
lattice data with all the pion masses less than $\sim 400$~MeV, to
properly map out the chiral regime.
This is a very exciting time, as we are
beginning to see lattice data of baryonic observables with $m_\pi \sim
330$~MeV, e.g.~\cite{Renner:2004ck}.
If not the pions accessible today, then the ones in the near future
will be light enough that one can apply chiral perturbation
theory to observables calculated from lattice QCD.
We anticipate that lattice QCD in conjunction
with \CPT\ and \PQCPT\ will soon be making predictions of nucleon
observables with theoretical errors at the 20\%
level~\cite{Beane:2004ks}.%
\footnote{We mean that the \CPT\ and \PQCPT\ expressions
  will be fit only to lattice calculations but not experimental
  values.  These expressions will then be used to extrapolate down to the
  physical quark masses, make predictions of experimental
  observables, and ultimately predictions of observables not
  currently accessible via experiment.}

Calculation of the baryonic mass spectrum from
lattice QCD data in the chiral regime will be a first step in
understanding nuclear physics from first principles.  To this end, we
determine the masses of the nucleons and delta-resonances 
to next-to-next-to-leading order (NNLO) in two-flavor non-degenerate
\CPT.  We highlight the mass splittings among the nucleons and deltas,
as they provide a way to determine some of the LECs entering at
NNLO.  This is because strong isospin breaking in the baryon spectrum
receives the first non-linear quark mass dependence at NNLO.
In the appendix we calculate the masses of the nucleons and
deltas in \PQCPT, with two non-degenerate valence and sea quarks.
These expressions are necessary to extrapolate partially quenched lattice data of the
nucleon and delta masses to the physical values of the quark masses.

%
%
%
%
%
%
%	SU(2) Masses
%
%
%
%
%
%

\section{Masses in \CPT} \label{CPT}

In this section, we calculate the masses of the nucleons and deltas 
in the non-degenerate SU($2$) flavor group. We start by reviewing the
chiral Lagrangian for mesons and baryons. Next we calculate the masses of the nucleons
and finally the masses of the deltas. The masses of the nucleons and 
deltas have been investigated considerably in 
\CPT, see e.g. \cite{Gasser:1980sb,Jenkins:1992ts,Bernard:1993nj,Lebed:1994yu,Lebed:1994gt,Banerjee:1995bk, 
  Borasoy:1997bx, Becher:1999he, Frink:2004ic, Walker-Loud:2004hf,
  Tiburzi:2004rh, Lehnhart:2004vi}.
In Appendix~\ref{ap:PQCPT}, we take up the partially quenched calculation.

%
%
%
%
%
%
%	Lagrangian
%
%
%
%
%
%

\subsection{\CPT\ Lagrangian}

For two flavors of massless quarks, the QCD Lagrangian exhibits a
chiral symmetry, SU$(2)_L \otimes$SU$(2)_R \otimes$U$(1)_V$. 
This symmetry is spontaneously broken by the vacuum down to SU$(2)_V \otimes$U$(1)_V$.
\CPT\ is the effective theory generated by expanding about the physical vacuum state
of QCD. Without the source of explicit chiral symmetry breaking generated by 
the quark mass term, the pions would appear as the Goldstone bosons of
this symmetry breaking.  In reality the up and down quark masses are not zero
but are quite small compared to the scale of chiral symmetry breaking. Thus the 
pions appear in nature as pseudo-Goldstone bosons.

To build the effective theory, we collect the pseudo-Goldstone bosons
in an exponential matrix 
\begin{equation}
  \S = 
    {\rm exp} \left( \frac{2\ i\ \Phi}{f} \right) = \xi ^2 \ ,\quad  
      \Phi  = 
        \begin{pmatrix}
          \frac{1}{\sqrt{2}}\pi^0 & \pi^+ \\
            \pi^- & -\frac{1}{\sqrt{2}}\pi^0 
        \end{pmatrix}.
\label{eq:Sigma}
\end{equation}
The pion decay constant $f$ is $132$~MeV in the above conventions.
The effective Lagrangian describing the dynamics of the pions at
leading order in $\chi$PT is \cite{Gasser:1983yg}
\begin{eqnarray}
  \mathfrak{L}  =\ 
    \frac{f^2}{8} \tr \left( \partial^\mu \S ^\dagger
      \partial_\mu \S \right)
     + \l \, \tr \left(m_q^\dagger  \S  + m_q \S^\dagger \right).
\label{eqn:Lmeson}
\end{eqnarray}
Above the quark mass matrix $m_Q$ is given by
\begin{equation}
m_q =  \diag (m_u, m_d).
\end{equation}
Expanding the Lagrangian Eq.~\eqref{eqn:Lmeson} one finds that to
leading order the pions are canonically normalized with their masses
given by
\begin{equation}
m^2_{\pi} = \frac{4 \lambda}{f^2} (m_u + m_d) 
\label{eq:mesonmass}.
\end{equation}

The baryons are three quark states in the spectrum of QCD. To include the 
lowest-lying spin-$\frac{1}{2}$ and spin-$\frac{3}{2}$ baryons into \CPT, 
we use heavy baryon \CPT\ 
(HB\CPT)~%
\cite{Jenkins:1991ne,Jenkins:1990jv,Jenkins:1991es,%
Jenkins:1992ts,Bernard:1992qa,Bernard:1995dp}. 
As is well known, this approach leads to a consistent power-counting scheme
for including the baryons.  In SU$(2)$, there is a doublet $N$ of 
spin-$\frac{1}{2}$ nucleons
\begin{equation}
N = 
\begin{pmatrix}
p \\
n
\end{pmatrix}
,\end{equation}
and a quartet of spin-$\frac{3}{2}$ deltas. These latter states are contained 
in the completely symmetric tensor $T^{ijk}$, where

\begin{equation}
T^{111} = \D^{++}, \qquad T^{112} = \frac{1}{\sqrt{3}} \D^+, \qquad T^{122} = \frac{1}{\sqrt{3}} \D^0, 
\qquad \text{and} \qquad T^{222} = \D^-
.\end{equation}

The free Lagrangian for nucleons and deltas to leading order in HB\CPT\ can be written as

\begin{eqnarray}
\mathfrak{L} 
&=& 
i \ol N \, v \cdot D \, N + 2 \a_M \ol N \, \c{M} \, N + 2 \sigma_M \ol N N \, \tr (\c{M})  
\notag \\
&& 
- i \ol T {}^\mu \, v \cdot D \, T_\mu + \D \, \ol T {}^\mu T_\mu + 2 \gamma_M \ol T {}^\mu \, \c{M} \, T_\mu 
- 2 \ol \sigma_M \ol T {}^\mu T_\mu \, \tr (\c{M})
.\end{eqnarray}
Above, $\D$ is the leading-order mass splitting between the nucleons
and deltas in the chiral limit. As a dimensionful parameter it must be
treated in the power counting and we treat $\D \sim \c{O}(q)$~\cite{Hemmert:1996xg,Hemmert:1997ye}. The
chirally covariant derivative $D^\mu$ acts on the nucleon field as

\begin{equation}
(D^\mu N)_i =  \partial^\mu N_i + (V^\mu)_{i}{}^{i'} N_{i'}
,\end{equation} 
and on the delta field as
\begin{equation}
(D^\mu T)_{ijk} = \partial^\mu T_{ijk} + (V^\mu)_{i}{}^{i'} T_{i'jk} + (V^\mu)_{j}{}^{j'} T_{ij'k} + (V^\mu)_{k}{}^{k'} T_{ijk'}
.\end{equation}
The vector and axial-vector meson fields are given by
\begin{equation}
  V_\mu =
    \frac{1}{2} ( \xi\partial_\mu\xi^\dagger 
      + \xi^\dagger\partial_\mu\xi ), \quad  
  A_\mu  =
    \frac{i}{2} ( \xi\partial_\mu\xi^\dagger 
      - \xi^\dagger\partial_\mu\xi )\, ,
\end{equation}
and finally the spurion mass field 
\begin{equation}
  \c{M}  = \frac{1}{2}\left( \xi^\dagger m_q \xi^\dagger + \xi m_q^\dagger
  \xi \right) 
.\end{equation}

The Lagrange density describing the interactions of the nucleons and deltas
with the pions at leading order can be written as
\begin{eqnarray}
\mathfrak{L} 
&=&
2 g_A \, \ol N \, S \cdot A \, N 
+ g_{\D N} \left( \ol T {}^\mu A_\mu N + \ol N A^\mu T_\mu \right)
+ 2 g_{\D \D} \ol T {}^\mu \, S \cdot A \, T_\mu
\label{eq:MBT},\end{eqnarray}
with $S^\mu$ as the covariant spin operator. 
Additionally we need higher-order terms to calculate masses at NNLO. 
The first set of higher-order terms are all operators that are constrained
by reparameterization 
invariance~\cite{Luke:1992cs,Manohar:2000dt}. 
These terms must be included to insure the Lorentz invariance of HB\CPT\ 
at $\c{O}(1/M_B)$\footnote{For an earlier, alternate approach for the nucleons, see, e.g., Ref.~\cite{Bernard:1992qa}.}, 
where $M_B \sim \Lambda_\chi$ is the average nucleon mass in the chiral limit. 
As a result of reparameterization invariance, these operators have fixed coefficients
and explicitly these operators are
\begin{eqnarray}
\mathfrak{L} 
&=& 
- \ol N  \frac{D_\perp^2}{2 M_B} N 
+ \Tbar {}^\mu \frac{D_\perp^2}{2 M_B} T_\mu 
+ g_A 
\left( 
\ol N \frac{i \loarrow D \cdot S}{M_B} v \cdot A \, N 
- 
\ol N \, v \cdot A \frac{ S \cdot i \roarrow D }{M_B} N
\right)
\notag \\
&&+ 
g_{\D \D} \left( \Tbar {}^\mu \frac{i \loarrow D \cdot S}{M_B}
       \vit \cdot A \, T_\mu - \Tbar {}^\mu \, \vit\cdot A \frac{S\cdot i \roarrow D}
       {M_B} T_\mu \right)
\label{eq:fixed}
,\end{eqnarray}
where $D_\perp^2 = D^2 - (\vit \cdot D)^2$.

Next there are three sets of higher-dimensional operators whose
coefficients are not pre-determined. The first set makes contributions
to the mass of the nucleons at NNLO. The operators in this set are
\begin{eqnarray}
	\mathfrak{L}  &=&
		\frac{1}{4 \pi f} \bigg\{
			b_1^M \ol N  \, \mathcal{M}_+^2  N
			+ b_5^M \ol N N \, \tr ( \mathcal{M}_+^2 )
			+ b_6^M \ol N \, \c{M} N \, \tr (\c{M}) 
			+ b_8^M \ol N N \, [\tr (\c{M})]^2 \notag \\
		&&\qquad\quad
		 	+ b^A \ol N N \, \tr ( A \cdot A )  
			+ b^{vA} \ol N N \, \tr ( v \cdot A \, v \cdot A ) 
		\bigg\}. \notag \\
\label{eq:NHDO} 
\end{eqnarray}
The LECs $b_i^M$, $b^A$, and $b^{vA}$ are all dimensionless. 
The choice in numbering the coefficients was made to be consistent
with~\cite{Walker-Loud:2004hf}, as will become clear in Appendix~\ref{ap:PQCPT}, where we 
address the partially quenched calculation.
The second set of higher-dimensional operators with undetermined coefficients 
consists of operators that contribute to the delta masses at NNLO. 
These operators are
\begin{eqnarray}\label{eq:DHDO}
	\mathfrak{L} &=& 
		\frac{1}{4 \pi f} \bigg\{
			t_2^A \, \ol T {}^{kji}_\mu (A_\nu)_{i}{}^{i'} (A^\nu)_{j}{}^{j'} T^\mu_{i'j'k} 
			+ t_3^A \, \left( \ol T_\mu T^\mu \right) \tr ( A_\nu A^\nu ) \notag \\
		&&\qquad\quad
			+ t_2^{\tilde{A}} \, \ol T {}^{kji}_\mu (A^\mu)_{i}{}^{i'} (A_\nu)_{j}{}^{j'} T^\nu_{i'j'k} 
			+ t_3^{\tilde{A}} \, \ol T_\mu T^\nu \tr ( A^\mu A_\nu ) \notag \\
		&&\qquad\quad
			+ t_2^{vA} \, \ol T {}^{kji}_\mu (v \cdot A)_{i}{}^{i'} (v \cdot A)_{j}{}^{j'} T^\mu_{i'j'k} 
			+ t_3^{vA} \, \ol T_\mu T^\mu  \tr ( v \cdot A \, v \cdot A ) \notag \\
		&&\qquad\quad
			+ t_1^M \, \ol T {}^{kji}_\mu (\c{M} \c{M})_{i}{}^{i'} T^\mu_{i'jk} 
			+ t_2^M  \, \ol T {}^{kji}_\mu (\c{M})_{i}{}^{i'} (\c{M})_{j}{}^{j'} T^\mu_{i'j'k} 
			+ t_3^M  \, \ol T_\mu T^\mu \tr (\c{M} \c{M}) \notag \\
		&&\qquad\quad
			+ t_4^M \, \left( \ol T_\mu \c{M} T^\mu \right) \tr (\c{M})
			+ t_5^M  \, \ol T_\mu T^\mu  [ \tr(\c{M}) ]^2
		\bigg\}
\label{eq:hdo},
\end{eqnarray}
and all of the LECs $t_i^M$, $t_i^A$, $t_i^{vA}$, and $t_i^{\tilde{A}}$ are dimensionless.

The last set of higher-dimensional operators with undetermined coefficients all involve 
the mass-splitting parameter $\D$,  which is a singlet under chiral transformations. 
Including the spin-$\frac{3}{2}$ fields in \CPT\ thus 
requires the addition of operators involving powers of $\D / \Lambda_\chi$. We shall not write out such 
operators explicitly. To account for the effects of these operators, all LECs in the calculation
must be treated as arbitrary polynomial functions of $\D / \Lambda_\chi$ and expanded out to the
required order. For example
\begin{eqnarray}\label{eq:DeltaExpnsn}
 \a_M \rightarrow \a_M \left(\frac{\D}{\L_\chi}\right)
      &=& \a_M \left( 1+\a_1 \, \frac{\D}{\L_\chi} 
        + \a_2 \, \frac{\D^2}{\L_\chi^2} \right),
        \notag \\
 \gm \rightarrow \gm\left(\frac{\D}{\L_\chi}\right)
      &=& \gm\left(1+\gamma_1 \, \frac{\D}{\L_\chi} 
        + \gamma_2 \, \frac{\D^2}{\L_\chi^2} 
        \right), \notag \\
 g_A \rightarrow g_A\left(\frac{\D}{\L_\chi}\right)
      &=& g_A \left(1 + g_1 \, \frac{\D}{\L_\chi} 
        \right)\, .
\end{eqnarray}
Determination of these LECs
requires the ability to tune the parameter $\D$
and for this reason we do not keep such operators explicitly.  
Furthermore at this order the nucleon mass, $M_0$, 
and the nucleon-delta mass splitting in the chiral limit are also functions of $\D/\Lambda_\chi$.
This variation, however, can be absorbed into a redefinition of the parameters $M_0$, and $\Delta$ 
(including contributions from loop graphs). We use these value for $M_0$ and $\D$ everywhere.  Differences resulting from using this value of $\D$ in expressions for loop graphs are beyond the order we work.

%
%
%
%
%
%
%	Nucleons
%
%
%
%
%
%
\subsection{Nucleons}

The mass of the $i^{th}$ nucleon in the chiral expansion can be written as
\begin{eqnarray}\label{eq:NucMassExp}
     M_{B_i} = M_0 \left(\D \right) -  M_{B_i}^{(1)}\left(\mu, \D \right)
                - M_{B_i}^{(3/2)}\left(\mu, \D \right)
                - M_{B_i}^{(2)}\left(\mu, \D \right) + \ldots
\label{eq:Bmassexp}
\end{eqnarray}
Here, $\mu$ is the renormalization scale and $\D$ is the quark mass independent delta-nucleon mass splitting.  $M_0 \left(\D \right)$ is the renormalized nucleon
mass in the chiral limit, is independent of $\mu$, of $m_q$ and also of the $B_i$.  $M_{B_i}^{(n)}$ is the contribution to the $i^{th}$ nucleon of the order $m_q^{(n)}$.\footnote{%
The renormalization scale dependence appears at each order because we are 
treating the LECs as polynomial functions of $\Delta$. If we instead
treat this dependence explicitly and expand the nucleon mass in powers of 
$q$, where $\Delta \sim q$ and $ m_q \sim q^2$, the mass then takes the form
\begin{eqnarray}
M_{B_i} 
= m_0  -  m_{B_i}^{(2)} - m_{B_i}^{(3)} - m_{B_i}^{(4)} + \ldots
\notag,\end{eqnarray}
with $m_{B_i}^{(n)}$ as the renormalization scale independent mass contribution 
to the $i^{\text{th}}$ nucleon strictly of the order $q^n$. The parameter $m_0$ is
nucleon mass in the chiral limit, but only to leading order in $\Delta$, i.e. $m_0 = M_0(\D=0)$.
}
Throughout this work, we use dimensional 
regularization with a modified minimal subtraction ($\ol{\rm   MS}$) 
scheme, where we have consistently subtracted terms proportional to 
\begin{equation}
\frac{1}{\varepsilon} - \gamma_E + 1 + \log 4 \pi
\notag 
.\end{equation}

The diagrams relevant to calculate the nucleon mass to NNLO 
are depicted in~\cite{Walker-Loud:2004hf}. 
We find that the leading-order contributions to the nucleon masses are
\begin{equation}\label{eq:LONmass}
  M^{(1)}_B 
    = 2 \a_M m_B 
     + 2 \sigma_M \, \tr (m_q),
\end{equation}
where
\begin{equation}\label{eq:mB}
  m_B = \left\{
    \begin{array}{lc}
      m_u, & B=p \\
      m_d, & B=n
    \end{array}\right. .
\end{equation}

The next-to-leading contributions are
\begin{equation}
  M^{(3/2)}_B = 
    \frac{3}{16 \pi f^2} g_A^2  m_\pi^3 
    + \frac{8 g_{\D N}^2}{3 (4 \pi f)^2} \c{F} (m_\pi,\D,\mu)
,\label{eq:MBNLO}
\end{equation}
where the function $\c{F}$ is defined by
\begin{eqnarray}
  \c{F} (m,\d,\mu) &=& (m^2 - \d^2) 
    \left[ \sqrt{\d^2 - m^2}
      \log \left( 
        \frac{\d - \sqrt{\d^2 - m^2 + i \varepsilon}}
        {\d + \sqrt{\d^2 - m^2 + i \varepsilon}} \right)
      - \d \log \left( \frac{m^2}{\mu^2} \right)
    \right] \notag \\
    && \phantom{sp} - \frac{1}{2} \d\, m^2 \log \left( \frac{m^2}{\mu^2} \right)
                    - \d^3 \log \left( \frac{4 \d^2}{\mu^2} \right)
\label{eqn:F}
.\end{eqnarray}

Finally the next-to-next-to-leading contributions to the nucleon mass are
\begin{eqnarray}
M_B^{(2)} &=& ( Z_B - 1) M_B^{(1)} 
  + \frac{1}{4 \pi f} \left\{
    b_1^M \, (m_B)^2 + b_5^M \, \tr (m_q^2) + b_6^M \, m_B \, \tr(m_q)
    + b_8^M \, [\tr(m_q)]^2 \right\} \notag \\
&& 
  - \frac{1}{(4 \pi f)^2} C^B_\pi \c{L} (m_\pi,\mu) 
    - \frac{6 \sigma_M}{(4 \pi f)^2} \, \tr(m_q)\,  \c{L} (m_\pi,\mu) \notag \\
&& 
  + \frac{3\, b^A}{(4 \pi f)^3} \ol{\c{L}} (m_\pi, \mu) 
    + \frac{3\, b^{vA}}{4 (4 \pi f)^3} \left[ \ol{\c{L}} (m_\pi, \mu) - \frac{1}{2} m_\pi^4 \right] \notag \\
&& 
  + \frac{1}{(4 \pi f)^2} \frac{27 g_A^2}{16 M_B} 
     \left[ \ol{\c{L}} (m_\pi, \mu) + \frac{5}{6} m_\pi^4 \right]
    + \frac{1}{(4 \pi f)^2} \frac{5 g_{\D N}^2}{2 M_B}
     \left[ \ol{\c{L}} (m_\pi, \mu) + \frac{9}{10} m_\pi^4 \right] \notag \\
&& 
  + \frac{ 9 g_A^2 \sigma_M }{ (4 \pi f)^2} \tr (m_q) 
     \left[ \c{L} (m_\pi, \mu) + \frac{2}{3} m_\pi^2 \right] 
    + \frac{8 g_{\D N}^2 \ol \sigma_M }{(4 \pi f)^2} \tr(m_q)
     \left[ \c{J} (m_\pi, \D, \mu) + m_\pi^2 \right] \notag \\
&& 
  + \frac{3g_A^2}{ (4 \pi f)^2} F^B_\pi \left[ \c{L} (m_\pi, \mu) + \frac{2}{3} m_\pi^2 \right]
     - \frac{2 g_{\D N}^2}{(4 \pi f)^2} \gamma_M G^B_\pi
     \left[ \c{J} (m_\pi, \D, \mu) + m_\pi^2 \right] 
.\label{e:MBNNLO} \end{eqnarray}
Here we have used $(m_B)^2$ which is the square of the tree-level
coefficients $m_B$ appearing in Eq.~\eqref{eq:mB}.
Above the wavefunction renormalization $Z_B$ is given by
\begin{equation}
  Z_B - 1 = 
     - \frac{9 g_A^2}{2 (4 \pi f )^2} \left[  \c{L}(m_\pi , \mu) +
       \frac{2}{3} m_\pi^2 \right]
     - \frac{4 g_{\D N}^2}{(4 \pi f)^2 } \left[ \c{J}(m_\pi, \D, \mu) +
       m_\pi^2 \right]
.\label{e:ZB} 
\end{equation}

The coefficients in the NNLO contribution, namely $C_\pi^B$, $F_\pi^B$, and $G_\pi^B$, are given in 
Table~\ref{t:NQCD-C} and depend on whether $B = p$ or $B = n$.
\begin{table}
\caption{The coefficients $C_\pi^B$, $F_\pi^B$, and $G_\pi^B$ in \CPT. Coefficients are
listed for the nucleons.}
%\begin{ruledtabular}
\begin{tabular}{l | c c c }
 & $\quad \quad \quad \quad C_\pi^B \quad \quad \quad \quad $  
 & $\quad \quad \quad \quad F_\pi^B \quad \quad \quad \quad$ 
 & $\quad \quad \quad \quad G_\pi^B  \quad \quad \quad \quad$ \\
\hline
$p$
 & $2 \a_M (2m_u +m_d)$ 
 & $\a_M (m_u + 2 m_d)$ 
 & $\frac{4}{9} ( 7 m_u + 2 m_d)$    \\
$n$
 & $2 \a_M (m_u + 2m_d)$ 
 & $\a_M (2 m_u + m_d)$ 
 & $\frac{4}{9} ( 2 m_u + 7 m_d)$    \\
 \end{tabular}
%\end{ruledtabular}
\label{t:NQCD-C}
\end{table}
The equations above, Eq.~\eqref{e:MBNNLO} and Eq.~\eqref{e:ZB}, also
employ abbreviations for non-analytic functions arising from 
loop contributions. These functions are
\begin{align}
\c{L} (m,\mu) &= m^2 \log \frac{m^2}{\mu^2}, \\
\ol{\c{L}} (m,\mu) &= m^4 \log \frac{m^2}{\mu^2}, \\
\c{J} (m,\d,\mu) &= (m^2 - 2 \d^2) \log \frac{m^2}{\mu^2} + 2 \d \sqrt{\d^2 - m^2} \log
\left(
\frac{\d - \sqrt{\d^2 - m^2 + i \varepsilon}}{\d + \sqrt{\d^2 - m^2 + i \varepsilon}}
\right) + 2 \d^2 \log \left( \frac{4 \d^2}{\mu^2} \right)
.\end{align}

The expressions we have derived in this section, as well as those
throughout this work, are functions of the quark masses;
e.g. $m_\pi$ above is 
merely a replacement for the combination of quark masses given in
Eq.~\eqref{eq:mesonmass}.  These expressions, thus require the
lattice practitioner to determine the quark masses.  
In $SU(2)$ \CPT\ away from the isospin limit, there are no equivalent
expressions in terms of the meson masses, as one cannot independently associate 
$m_u$ and $m_d$ to the pion masses, which are
degenerate to the order we are working.  One can only equate the
average value, $\frac{1}{2} (m_u +m_d)$, with the pion mass.
Therefore one cannot plot the 
baryon masses as a function of the physical meson masses, unless one
works in the isospin limit, $m_u = m_d$.  This is unlike the case in
the isospin limit of $SU(3)$, where there are only two independent
quark masses, but three independent meson masses, and one can always
convert from a quark mass expansion to a meson mass expansion via the
Gell-Mann--Okubo relation.\footnote{%
This problem can be avoided in $SU(4|2)$ \PQCPT, as there are more
independent meson masses than independent quark masses, so one can
algebraically convert from the quark mass expansion to the
meson mass expansion.  For example, to leading order,
$m_u = \frac{f^2}{8 \lambda} \left(m_\pi^2 -m^2_{jd} +m^2_{ju} \right)$.
This would require one to know the mass of the mesons made of one
valence and one sea quark.  But if one is interested in the
non-isospin $SU(2)$ limit of the \PQCPT\ expressions, the problem is unavoidable.}
Lastly we remark that if one is using \CPT\ to determine the quark
masses from meson masses that are determined on the lattice, one must
use the one-loop expression in \CPT, else one looses contributions to
baryon masses that are of NNLO.  Thus the quark mass expansion is the one to
use in $SU(2)$ \CPT, at least away from the
isospin limit.

%
%
%
%
%
%
%	Deltas
%
%
%
%
%
%
\subsection{Deltas}

The mass of the $i^{th}$ delta in the chiral expansion can be written as
\begin{eqnarray}
     M_{T_i} = M_0 \left(\D \right) + \D +  M_{T_i}^{(1)}\left(\mu, \D \right)
                + M_{T_i}^{(3/2)}\left(\mu, \D \right)
                + M_{T_i}^{(2)}\left(\mu, \D \right) + \ldots
\label{eq:Tmassexp}
\end{eqnarray}
Here, $M_0 \left(\D \right)$ is the renormalized nucleon mass 
in the chiral limit from Eq.~\eqref{eq:NucMassExp}, and $\D$ is the renormalized nucleon-delta mass splitting in the chiral limit. 
Both of these quantities are independent of $m_q$ and also of the $T_i$.
$M_{T_i}^{(n)}$ is the contribution to the 
$i^{th}$ decuplet baryon of the order $m_q^{(n)}$, and $\mu$ is the
renormalization scale.\footnote{%  
As with the nucleon masses, the renormalization scale appears in each term 
contributing to the delta masses because we implicitly
treat the LECs as polynomial functions of the mass parameter $\Delta$. 
If we expand out the LECs, the $\mu$-dependence disappears at each order.}%

The diagrams relevant to calculate the delta masses
to NNLO are depicted in~\cite{Tiburzi:2004rh}. 
We find that the leading-order contributions to the delta masses are
\begin{equation}
M^{(1)}_T = \frac{2}{3} \, \gamma_M \, m_T  - 2 \ol \sigma_M \, \tr (m_q),
\label{eq:MLO}
\end{equation}
where the tree-level coefficients $m_T$ are given in Table \ref{t:mT} for the deltas $T$. 
\begin{table}
\caption{The tree-level coefficients in \CPT. The coefficients $m_T$,
  $(m^2)_T$, and $(mm')_T$ are listed for the deltas $T$.}
%\begin{ruledtabular}
\begin{tabular}{l | c c c }
 & $m_T$ & $ \phantom{spac} (m^2)_T \phantom{spac}$ & $(mm')_T$ \\
\hline
$\D^{++}$       
             &  $3 m_u$ & $3 m_u^2$  & $3m_u^2$  \\
$\D^+$ 
             &  $2 m_u + m_d$ & $2 m_u^2 + m_d^2$  & $m_u^2 + 2 m_u m_d$  \\
$\D^0$    
             &  $m_u + 2 m_d$ & $m_u^2 + 2 m_d^2$  & $2 m_u m_d + m_d^2$  \\
$\D^-$ 
             &  $3 m_d$ & $3 m_d^2$  & $3 m_d^2$  \\
\end{tabular}
%\end{ruledtabular}
\label{t:mT}
\end{table}
The next-to-leading order contributions are
\begin{equation}
M^{(3/2)}_T = -\frac{25 g_{\D\D}^2}{432 \pi f^2} \, m_\pi^3 
- \frac{2 g_{\D N}^2}{3(4 \pi f)^2} \, \c{F} (m_\pi,-\D,\mu)
.\label{eq:MNLO}
\end{equation}

Finally the next-to-next-to-leading contributions to the delta masses are
\begin{eqnarray}
M_T^{(2)} &=& (Z_T - 1) M_T^{(1)} \notag \\ 
&& +
    \frac{1}{4 \pi f} \left\{ \frac{1}{3} t_1^M \, (m^2)_T 
      + \frac{1}{3} t_2^M \, (m m')_T 
      + t_3^M \, \tr(m_q^2) 
      + \frac{1}{3} t_4^M \,  m_T  \, \tr (m_q)
      + t_5^M \, [\tr (m_q)]^2 \right\} \notag \\
&& -
    \frac{2 \,\gamma_M}{(4 \pi f)^2}  C_\pi^T \, \c{L}(m_\pi,\mu) 
	+ \frac{6 \, \ol \sigma_M}{(4 \pi f)^2} \, \tr(m_q) \,
		\c{L}(m_\pi,\mu)
	+ \frac{1}{(4 \pi f )^3} \left( 
		\frac{1}{2} t^A_2  + 3 t^A_3 \right)
		\ol{\c{L}} (m_\pi, \mu) \notag \\
	&&	+\frac{1}{4 (4 \pi f )^3} \left[ 
		\frac{1}{2} (t^{\tilde{A}}_2 + t_2^{vA}) 
		+ 3 (t^{\tilde{A}}_3  + t_3^{vA}) \right]
			\left[ \ol{\c{L}} (m_\pi, \mu)
			- \frac{1}{2} m_\pi^4 \right] \notag \\
&& -
    \frac{25 g_{\D\D}^2}{48 (4\pi f)^2 M_B} 
     \left[
       \ol{\c{L}}(m_\pi,\mu) + \frac{19}{10} m_\pi^4 \right]
      - \frac{5 g_{\D N}^2}{8 (4 \pi f)^2 M_B}  
       \left[
         \ol{\c{L}}(m_\pi,\mu) - \frac{1}{10} m_\pi^4 \right] \notag \\
&& -  
    \frac{25 g_{\D\D}^2 {\ol \sigma_M}}{9 (4 \pi f)^2} \tr(m_Q)
     \left[ \c{L}(m_\pi,\mu) + \frac{26}{15} m_\pi^4 \right]
    - \frac{2 g_{\D N}^2 \sigma_M}{(4 \pi f)^2} \tr(m_Q)
     \c{J} (m_\pi,-\D,\mu)\notag \\
&& +
    \frac{10 g_{\D\D}^2 \gamma_M}{9 (4 \pi f)^2} 
      F_\pi^T \left[ \c{L}(m_\pi,\mu) + \frac{26}{15} m_\pi^2 \right]
      - \frac{3 g_{\D N}^2 }{2 (4 \pi f)^2} G_\pi^T  \, \c{J} (m_\pi,-\D,\mu)
\label{eq:MNNLO}
.\end{eqnarray}
Here we have used $(m^2)_T$, $(m m')_T$ to label the tree-level coefficients that
appear in Table \ref{t:mT}. 
Above the wavefunction renormalization $Z_T$ is given by
\begin{equation}
Z_T - 1 = - \frac{25 g_{\D\D}^2}{18 (4 \pi f)^2} 
           \left[ \c{L}(m_\pi,\mu) + \frac{26}{15} m_\pi^2 \right]
          - \frac{g_{\D N}^2}{(4 \pi f)^2} \c{J} (m_\pi,-\D,\mu)
\label{eq:Z}
.\end{equation}
The coefficients in the NNLO contribution, namely $C_\pi^T$, $F_\pi^T$, and $G_\pi^T$, are given in 
Table~\ref{t:QCD-C} and depend on the delta state $T$.

\begin{table}
\caption{The coefficients $C_\pi^T$, $F_\pi^T$, and $G_\pi^T$ in \CPT. Coefficients are
listed for the delta states $T$.}
%\begin{ruledtabular}
\begin{tabular}{l | c c c }
 & $\quad \quad \quad C_\pi^T \quad \quad \quad $  
 & $\quad \quad \quad F_\pi^T \quad \quad \quad$ 
 & $\quad \quad \quad G_\pi^T  \quad  $ \\
\hline
$\D^{++}$       
 & $2 m_u + m_d$  
 & $\frac{13}{6} m_u + \frac{1}{3} m_d$ 
 & $\frac{4}{3} \a_M m_u$    \\
$\D^+$ 
 & $\frac{5}{3} m_u + \frac{4}{3} m_d$ 
 & $\frac{14}{9} m_u + \frac{17}{18} m_d$ 
 & $\frac{4}{9} \a_M ( 2 m_u + m_d )$\\
$\D^0$    
 & $\frac{4}{3} m_u + \frac{5}{3} m_d$  
 & $\frac{17}{18} m_u + \frac{14}{9} m_d$ 
 & $\frac{4}{9} \a_M (m_u + 2 m_d)$\\
$\D^-$ 
 & $m_u + 2 m_d$ 
 & $\frac{1}{3} m_u + \frac{13}{6} m_d$ 
 & $\frac{4}{3} \a_M m_d$  \\
 \end{tabular}
%\end{ruledtabular}
\label{t:QCD-C}
\end{table}

%
%
%
%
%
%
%	mass Splittings
%
%
%
%
%
%
\section{Mass Splittings}\label{MSplit}

Having derived the nucleon and delta masses to NNLO in the chiral
expansion, we now focus on the mass splittings between these states.
To begin, we consider the nucleon mass splitting, which to our knowledge was first theoretically addressed in~\cite{Gasser:1974wd}. The degeneracy
between the proton and neutron is broken by leading-order effects
in the chiral theory, see Eq.~(\ref{eq:LONmass}). Beyond this order
pion loops contribute, but to the order we are working, all the pions
are degenerate, even away from the isospin limit.  Thus the
NLO contributions to the neutron and proton masses are the same,
see Eq.~(\ref{eq:MBNLO}), and the mass splitting, $M_n
- M_p$, is given to NLO accuracy, entirely by the difference of the LO mass
contribution in Eq.~(\ref{eq:LONmass}), and is linear 
in $m_d - m_u$.   Any deviation from this linear mass splitting seen in
lattice simulations of the nucleon masses should be a signature of the NNLO
mass contributions and certain LECs that arise at this
order.  Additionally, the nucleon mass splitting can be enhanced from
that in nature by increasing the quark mass splitting on the lattice, $m_d - m_u$.
This enhancement, combined with the vanishing of the NLO contribution
to the mass splitting, provides us with a means of cleanly determining
the NNLO nucleon mass contributions and isolating certain LECs arising
at this order.  These effects are normally obscured by the NLO contributions.

%\subsection{Nucleon Mass Splittings}
We find the nucleon mass splitting is given to NNLO by

\begin{eqnarray}\label{eq:NmassSplit}
  M_n - M_p &=& 
    - 2\a_M (m_d - m_u) \nonumber\\
  &&+ (m_d - m_u) \bigg\{ \frac{m_\pi^2}{(4\pi f)^2} \bigg[
        8\left( g_A^2 \a_M + g_{\D N}^2 \left( \a_M +\frac{5}{9}\g_M
         \right) \right)
        - (b^M_1 +b^M_6)\frac{\pi f^3}{\l} \bigg] \nonumber\\
  &&\qquad\qquad\qquad
     + \frac{\c{L}(m_\pi,\mu)}{(4\pi f)^2}\ 2\a_M (6 g_A^2 +1)
     \nonumber\\
  &&\qquad\qquad\qquad
     + \frac{\c{J} (m_\pi,\D,\mu)}{(4\pi f)^2}\ 8 g_{\D N}^2
        \left( \a_M + \frac{5}{9} \g_M \right) \bigg\}.
\end{eqnarray}

An identical situation arises for the deltas, as their degeneracy
is broken at leading order in the chiral expansion, while the next
contribution to their splittings occurs at NNLO. Their mass
splittings are given by $\d M_T$, which stands for $M_{\D^-}
-M_{\D^0}$\ ,\ $M_{\D^0} -M_{\D^+}$ and $M_{\D^+} -M_{\D^{++}}$.  We find,

\begin{eqnarray}\label{eq:DmassSplit}
  \d M_T &=& \frac{2}{3}\g_M (m_d -m_u) \nonumber\\
  &&+ (m_d -m_u) \bigg\{ \frac{m_\pi^2}{(4\pi f)^2} \bigg[
        \frac{\pi f^3}{3 \l}(t^M_1 +t^M_4) 
        -\frac{104}{243} g_{\D\D}^2 \g_M \bigg] \nonumber\\
  &&\qquad\qquad\qquad
    - \frac{\c{L}(m_\pi,\mu)}{(4\pi f)^2} \frac{20\g_M}{81} 
     \left( \frac{27}{10}+ g_{\D\D}^2 \right) \nonumber\\
  &&\qquad\qquad\qquad
    - \frac{\c{J} (m_\pi,-\D,\mu)}{(4\pi f)^2} \frac{2 g_{\D N}^2}{3} (\g_M
    +\a_M) \nonumber\\
  &&\qquad\qquad\qquad
    + \frac{1}{12 \pi f} t^M_2\ \d t^M  \bigg\},
\end{eqnarray}
where $\d t^M$ is given in Table~\ref{t:MassSplitViol}.  Considering
these splittings, there are a few things to note.  We know from
experiment that the neutron 
is more massive than the proton, and we expect that this is mostly due
to $m_d > m_u$.  Similarly, we expect the deltas to follow a
similar pattern with $\D^-$ being more massive than $\D^0$, and so on,
although this is presently undetermined experimentally.  From these
expectations and the expressions above, it is expected that $\a_M$ is
negative and $\g_M$ is positive.  Also, the delta masses, to a good
approximation are expected to follow an equal spacing rule.  In HB\CPT,
this rule is first violated at NNLO, and even then, only by the
operator in Eq.~(\ref{eq:DHDO}) with coefficient $t^M_2$.  This
coefficient can be isolated by taking successive differences.

\begin{table}
\caption{The coefficient $\d t^M$, which encodes the violation of
  the delta equal spacing rule.}
\begin{tabular}{c | c }
   $\d M_T$
 & $\d t^M$\\
\hline
$M_{\D^-} -M_{\D^0}$    
 & $2m_d$\\
$M_{\D^0} -M_{\D^+}$ 
 & $m_u +m_d$\\ 
$M_{\D^+} -M_{\D^{++}}$       
 & $2 m_u$  
 \end{tabular}
\label{t:MassSplitViol}
\end{table}

Lastly we should comment that in nature isospin violation in the
baryon masses has another source of the same size as the NNLO chiral
effects, namely electromagnetic contributions.  In lattice QCD calculations one can turn off the electric charges of the quarks.  This is the scenario for which our calculations are applicable.  There have been a few lattice computations of electromagnetic contributions to hadronic masses~%
\cite{Duncan:1996xy,Duncan:1996be, Duncan:2004ys}.  For a recent discussion of the electromagnetic effects in hadrons, see Ref.~\cite{Gasser:2003hk}, and for a comprehensive phenomenological review, 
\cite{Miller:1990iz}.
%
%
%
%
%
%
%	Summary
%
%
%
%
%
%
\section{Summary}\label{Summary}
The expansion parameter for heavy baryon \CPT\ is $m_\pi / \L_\chi$ as opposed
to $m_\pi^2 / \L_\chi^2$ in the purely mesonic sector.  Therefore it
is crucial to calculate baryonic observables to at least NNLO to test
the convergence of the heavy baryon chiral expansion, as well as
reduce the theoretical error bars of these calculations.  To this
order, we are forced to introduce a large number of LECs [see 
Eq.(\ref{eq:NHDO}-\ref{eq:DHDO})], which must be determined to have any
predictive power, however, this is no small feat.  To NNLO, there are
more than ten LECs which must be fit to determine the nucleon masses [not
including those associated with the $\D / \L_\chi$ expansion, see
Eq.~(\ref{eq:DeltaExpnsn})], and there are only two nucleons.  Some of
these can be fit from observables besides the mass, but even so, not
all can be reliably determined.  Lattice QCD simulations provide us
with another means of determining the LECs, as we have the
freedom to vary the quark masses, providing a much larger data set
to fit these universal constants.  Today, these lattice simulations
are done with quark masses which are significantly larger than those in
nature.  It is therefore crucial to understand the quark mass
dependence of hadronic observables: in particular for nuclear physics,
those of the nucleons and deltas.  In this way, the LECs can be fit,
allowing the lattice data to be extrapolated from the valence and sea
quark masses used to their physical values, thus making rigorous
QCD predictions of these hadronic observables.

In this work we have calculated the masses of the nucleons and
the deltas in non-degenerate, two-flavor heavy baryon chiral
perturbation theory to $\c{O} (m_q^2)$.  One can obtain from the
authors a Mathematica
notebook containing the derived expressions in various
limits, including the partially-quenched calculations worked out in Appendix~\ref{ap:PQCPT}, relevant for current and future lattice data.
We have highlighted the
mass splittings of the nucleons and deltas as the $\c{O} (m_q^{3/2})$
contribution to these mass differences vanish, even away from 
the isospin limit.  To LO these mass differences are proportional to
$(m_d - m_u)$.  Any deviations from this linear behavior are a signature
of the NNLO mass contributions.  This provides a clean handle
to determine some of the LECs which contribute to the masses at $\c{O}
(m_q^2)$, which can otherwise be rather difficult to extract.

Another example of the utility of \CPT\ is to predict the width of the
deltas from the lattice.  As the deltas are resonances, one cannot
easily obtain their physical properties directly from real-valued 
Euclidean space simulations.  As the lattice pion masses are brought
down, one must deal with unstable particles.  Instead, one can work
with pion masses $m_\pi \gtrsim 300$ MeV for which the deltas are
stable and lattice calculations 
can be performed more simply than the unstable case (which also has
yet to be explored). In this regime, 
one still hopes the chiral expansion is valid,~\footnote{%
This is only a conservative estimate as finite volume effects will
modify the decay threshold of the deltas.}
and uses the \CPT\ expressions derived above, or the \PQCPT\
expressions derived in Appendix~\ref{ap:PQCPT}, to fit the
LECs and these consequently lead to predictions for the physical delta
resonances. For example, their decay widths can be found to NNLO from
knowing the pion mass and the constants $f$, $\lambda$, $\D$, $\a_M$,
$\sigma_M$, $\gamma_M$, $\ol \sigma_M$, and $g_{\D N}$.  For
simplicity the expression for the average delta width $\Im \text{m}
(\ol M_T)$ in terms of these parameters is 
%\begingroup
%\small
\begin{equation}
  \Im \text{m} (\ol M_T) = \left\{
    \begin{array}{lc}
      - \frac{g_{\D N}^2}{12 \pi f^2} \sqrt{\D^2 - m_\pi^2} 
        \bigg\{ \D^2 - m_\pi^2 %& \\
%        \qquad\qquad\qquad
        + \D \frac{f^2 m_\pi^2 }{24 \lambda}
          \left[ 3 \gamma_M  + 4  (\sigma_M - \ol \sigma_M )  + 2 \a_M
          \right]
        \bigg\}, & \text{for } 
           m_\pi < \D \\
      \qquad\qquad\qquad 0\, , & \text{for } 
           m_\pi > \D %\\
    \end{array}
    \right.
\end{equation}
%\endgroup

As a further application, the mass splitting between the nucleons and
deltas can also be found from the various LECs.  
For the real part of the difference between the average nucleon and average delta masses, we have\footnote{%
Here we have used the following replacements $A-G$ for particular combinations of the LECs: 
$A = \a_M + 2 (\sigma_M - \ol \sigma_M ) + \gamma_M$,
$B = \frac{3}{2} b_1^M + b_4^M + \frac{1}{2} t_1^M + t_3^M$,
$C = b_5^M + b_7^M + \frac{1}{3} t_2^M + \frac{1}{2} t_4^M + t_5^M$, 
$D = \frac{3}{2} b_1^A + 3 b_4^A + \frac{3}{8} b_1^{vA} + \frac{3}{4} b_4^{vA} + \frac{3}{2} t_1^A + \frac{1}{2} t_2^A + 3 t_3^A 
+ \frac{3}{8} (t_1^{\tilde{A}} + t_1^{vA}) + \frac{1}{8} (t_2^{\tilde{A}} + t_2^{vA}) + \frac{3}{4} (t_3^{\tilde{A}} + t_3^{vA})
$,
$E = \frac{27}{2} g_A^2 +  15 g_{\D N}^2 - \frac{25}{6} g_{\D \D}$, 
$F = - \frac{1}{8} \left[ \frac{3}{2} b_1^{vA} + 3 b_4^{vA} 
+ \frac{3}{2} (t_1^{\tilde{A}} + t_1^{vA}) + \frac{1}{2} (t_2^{\tilde{A}} + t_2^{vA}) + 3 (t_3^{\tilde{A}} + t_3^{vA})\right]
$, and
$G = \frac{45}{4} g_A^2 + \frac{37}{2} g_{\D N}^2 - \frac{95}{12} g_{\D \D}^2$.
}
\begin{eqnarray}
\Re \text{e} (\ol M_T - \ol M_B) 
&=& 
\Delta 
+ 
A \, \tr (m_q) 
+ 
B \, \tr (m_q^2) 
+ 
C \, [\tr(m_q)]^2 
\notag \\  
&& 
+ \frac{m_\pi^3}{16 \pi f^2} \left[ 3 g_A^2 - \frac{25}{27} g_{\D \D}^2 \right] 
+ \frac{5 g_{\D N}^2}{24 \pi^2 f^2} \c{F}(m_\pi, \D, \mu) \notag \\
&& 
+
\frac{A \, \tr(m_q)}{(4 \pi f)^2} 
\left\{
3  \c{L} (m_\pi, \mu) 
- 
5 g_{\D N}^2  
\left[ 
\c{J} (m_\pi, \D, \mu) 
+ 
\frac{4}{5} m_\pi^2 
\right]
\right\}
\notag \\
&& +
\frac{1}{(4 \pi f)^2} \ol{\c{L}}(m_\pi,\mu) 
\left[ 
\frac{1}{4 \pi f} D + \frac{1}{8 M_B} E
\right]
+ 
\frac{m_\pi^4}{(4 \pi f)^2}
\left[ 
\frac{1}{4 \pi f} F + \frac{1}{8 M_B} G
\right]. \notag \\
\end{eqnarray}
Once the LECs are known from the calculation of baryon masses, say,
one can use the chiral effective theory to calculate certain
contributions to $\pi N \to \pi N$ or $\pi \D \to \pi \D$ scattering,
for example, two processes which are complicated to explore
on the lattice.

The nucleons and deltas hold an important place in hadronic
physics. As the lightest spin-$\frac{1}{2}$ and spin-$\frac{3}{2}$
baryons, they have received considerable attention theoretically to
compare with a wealth of experimental data.  Soon lattice QCD
calculations will enable an understanding of the nucleon and delta
observables in terms of the quark masses.  A crucial first
step will be determining their masses from first principles. Current
and foreseeable lattice calculations will rely upon chiral
extrapolations.  Our expressions for the masses in \CPT\ and \PQCPT\
enable such extrapolations, and provide a means to access physics that
cannot be directly studied on current lattices.

\begin{acknowledgments}
We thank Martin Savage, Will Detmold and Matt Wingate for many useful
discussions.  This work was supported in part by the U.S. Department of
Energy under Grants No.~DE-FG03-97ER41014 (A.W.-L.) and
DE-FG02-96ER40945 (B.C.T.).
\end{acknowledgments}

%\newpage

\appendix

%
%
%
%
%
%
%	PQCPT
%
%
%
%
%
%
\section{Partially Quenched Chiral Perturbation Theory}\label{ap:PQCPT}
The study of PQQCD is motivated by lattice QCD simulations.  Presently
all lattice QCD simulations are done with quark masses that are
larger than those in nature, due to the computational cost of
simulating lighter quarks.  The valence quarks, those connected to
external fields, can be treated independently from the dynamical
sea-quarks that exist only in the closed quark
loops.  When the masses of the valence and sea quarks are taken to be
different, one has partially quenched QCD (PQQCD), and in the limit
that the valence and sea 
quark masses are equal, one has QCD.  To exclude contributions from
closed valence loops, one must include ghost quarks in the
effective theory of the lattice action,
which are identical to the valence
quarks, except that they have bosonic statistics.  Thus the valence
and ghost quark contributions to the PQQCD functional integral exactly
cancel, leaving only the sea-quark contributions.

As in QCD, one postulates that the vacuum of PQQCD spontaneously
breaks the (graded) chiral symmetry down to the vector subgroup,
and that one can perturb about the massless quark limit.  The emerging
low-energy effective theory of PQQCD is partially quenched \CPT\
(\PQCPT)~\cite{Bernard:1993sv,Sharpe:1997by,Golterman:1997st, 
  Sharpe:2000bc,Sharpe:2001fh,Sharpe:2003vy}.  Since PQQCD retains an
axial anomaly, the flavor singlet field can be integrated out and
there are no additional operators required that involve the singlet
field.  Moreover because the sea sector of PQQCD contains QCD, the
same LECs of \CPT\ appear in \PQCPT\ with the same numerical values.
Hence there is a means to determine \CPT\ parameters using PQQCD
lattice simulations. This hope has generated much activity in
calculating baryon masses in partially quenched
theories~\cite{Chen:2001yi,Beane:2002vq,Walker-Loud:2004hf,
  Tiburzi:2004rh,Tiburzi:2004kd}.

In PQQCD, the quark part of the Lagrangian is
\begin{equation}
\mathfrak{L} = \sum_{j,k=1}^6 \ol{Q}{}^{\hskip 0.2em j} \left(
  i\Dslash - m_Q \right)_j^{\hskip 0.3em k} Q_k
.\label{eq:pqqcdlag}
\end{equation}
This differs from the usual $SU(2)$ Lagrangian of QCD by the
inclusion of four extra quarks; two bosonic ghost quarks, ($\tilde u,
\tilde d$), and two fermionic sea quarks, ($j, l$), in addition
to the light physical quarks ($u, d$).  The
six quark fields transform in the fundamental representation of the
graded $SU(4|2)$ group.  They have been accommodated in the six-component vector
\begin{equation}
  Q^T = (u, d, j, l, \tilde{u}, \tilde{d})
.\end{equation}
The quark fields obey the graded equal-time commutation relation
\begin{equation}
Q^\a_i(\mathbf x) Q^{\beta \dagger}_j(\mathbf y) -
(-1)^{\eta_i \eta_j} Q^{\b \dagger}_j(\mathbf y) Q^\a_i(\mathbf x) =
\d^{\a \b} \d_{ij} \d^3 (\mathbf {x-y}),
\end{equation}
where $\a, \beta$ and $i,j$ are spin and flavor indices, respectively.
Analogous graded equal-time commutation relations can be written for
two $Q$'s and two $Q^\dagger$'s.  The grading factors
\begin{equation}
   \eta_k
   = \left\{ 
       \begin{array}{cl}
         1 & \text{for } k=1,2,3,4 \\
         0 & \text{for } k=5,6
       \end{array} 
     \right. ,
\end{equation}
take into account the different statistics 
of the quark fields of PQQCD.  The quark mass matrix of
$SU(4|2)$ is given by
\begin{equation}
  m_Q = \diag(m_u, m_d, m_j, m_l, m_u, m_d).
\end{equation}
QCD is recovered in the limit $m_j \rightarrow m_u$ and $m_l \rightarrow m_d$.

%
%
%
%
%
%
%	PQ lagrangian
%
%
%
%
%
%
\subsection{Partially Quenched $\chi$PT Lagrangian}

For massless quarks, the theory corresponding to the Lagrangian in
Eq. (\ref{eq:pqqcdlag}) has a 
graded $SU(4|2)_L \otimes SU(4|2)_R \otimes~U(1)_V$ symmetry which is
assumed to be spontaneously broken down to $SU(4|2)_V \otimes U(1)_V$
in analogy with QCD.  The effective low-energy theory obtained by
perturbing about the physical vacuum state of PQQCD is PQ$\chi$PT~%
\cite{Bernard:1993sv,Sharpe:1997by,Golterman:1997st, 
  Sharpe:2000bc,Sharpe:2001fh,Sharpe:2003vy}.
The emerging pseudo-Goldstone mesons have dynamics described 
at leading order in the chiral expansion by the Lagrangian
\begin{equation}
  \mathfrak{L} =
    \frac{f^2}{8} \str \left(
      \partial^\mu \S^\dagger \partial_\mu \S \right)
      + \l  \, \str \left( m_Q \S^\dagger + m_Q^\dagger \S \right)
\label{eq:pqbosons}
\end{equation}
where
\begin{equation}
  \S = \exp \left( \frac{2 i \Phi}{f} \right) = \xi^2
,\end{equation}
and the meson fields appear in
\begin{equation}
    \Phi =
    \begin{pmatrix}
      M & \chi^\dagger\\
      \chi & \tilde M\\
    \end{pmatrix}.
\end{equation}
The operation $\str( )$ in Eq. (\ref{eq:pqbosons}) is a graded flavor trace.  
$M$ and $\tilde M$ are
matrices containing bosonic mesons (with quantum numbers of $q \bar{q}$ pairs and 
$\tilde{q} \bar{\tilde{q}}$ pairs, respectively), while $\chi$ and $\chi^\dagger$
are matrices containing fermionic mesons (with quantum numbers of $\tilde q \bar{q}$
pairs and $q \bar{\tilde{q}}$ pairs, respectively).
The upper $2 \times 2$ block of $M$ contains the familiar
pions and the remaining components are mesons formed
with one or two sea quarks, see e.g.~\cite{Beane:2002vq}.
Expanding the Lagrangian Eq.~\eqref{eq:pqbosons} to leading order,
one finds that mesons with quark content $QQ'$ are canonically normalized
with their masses given by
\begin{equation}
m^2_{QQ'} = \frac{4 \lambda}{f^2} (m_Q + m_{Q'}) 
\label{eq:pqmesonmass}.
\end{equation}

The flavor singlet field is defined to be $\Phi_0 = {\rm str}( \Phi ) /
{\sqrt 2}$.  As PQQCD has a strong axial anomaly $U(1)_A$, 
the mass of the singlet field is on the order of the chiral symmetry breaking scale 
and has been integrated out of the theory~\cite{Sharpe:2001fh}.  
In this limit, however, the $\eta$
two-point correlation functions deviate from their form in \CPT.
For $a,b = u,d,j,l,\tilde u, \tilde d$, 
the leading-order $\eta_a \eta_b$ propagator with non-degenerate 
sea-quarks is
\begin{equation}
{\cal G}_{\eta_a \eta_b} =
        \frac{i \epsilon_a \delta_{ab}}{q^2 - m^2_{\eta_a} +i\epsilon}
        - \frac{i}{2} \frac{\epsilon_a \epsilon_b \left(q^2 - m^2_{jj}
            \right) \left( q^2 - m^2_{ll} \right)}
            {\left(q^2 - m^2_{\eta_a} +i\epsilon \right)
             \left(q^2 - m^2_{\eta_b} +i\epsilon \right)
             \left(q^2 - m^2_X +i\epsilon \right)}\, ,
\end{equation}
where
\begin{equation}
\epsilon_a = (-1)^{1+\eta_a}
.\end{equation}
The mass $m_{xy}$ is the mass of a meson composed of (anti)-quarks
of flavor $x$ and $y$, while the mass $m_X$ is defined as $m_X^2 =
\frac{1}{2}\left(m^2_{jj} + m^2_{ll}\right)$.  The flavor neutral  
propagator can be conveniently rewritten as
\begin{equation}
{\cal G}_{\eta_a \eta_b} =
         \e_a \d_{ab} P_a +
         \e_a \e_b {\cal H}_{ab}\left(P_a,P_b,P_X\right),
\end{equation}
where
\begin{eqnarray}
     P_a &=& \frac{i}{q^2 - m^2_{\eta_a} +i\e},\ 
     P_b = \frac{i}{q^2 - m^2_{\eta_b} +i\e},\ 
     P_X = \frac{i}{q^2 - m^2_X +i\e}, \,
\nonumber\\
\nonumber\\
\nonumber\\
     {\cal H}_{ab}\left(A,B,C\right) &=& 
           -\frac{1}{2}\left[
             \frac{\left( m^2_{jj}-m^2_{\eta_a}\right)
                   \left( m^2_{ll}-m^2_{\eta_a}\right)}
                  {\left( m^2_{\eta_a} - m^2_{\eta_b}\right)
                   \left( m^2_{\eta_a} - m^2_X\right)}
                 A
            -\frac{\left( m^2_{jj}-m^2_{\eta_b}\right)
                   \left( m^2_{ll}-m^2_{\eta_b}\right)}
                  {\left( m^2_{\eta_a} - m^2_{\eta_b}\right)
                   \left( m^2_{\eta_b} - m^2_X\right)}
                 B \right.\, 
\nonumber\\
&&\qquad\quad\left.
            +\frac{\left( m^2_X-m^2_{jj}\right)
                   \left( m^2_X-m^2_{ll}\right)}
                  {\left( m^2_X-m^2_{\eta_a}\right)
                   \left( m^2_X-m^2_{\eta_b}\right)}
                 C\ \right].
\label{eq:Hfunction}
\end{eqnarray}

For our calculation, the nucleons and deltas must also be included in
\PQCPT~\cite{Chen:2001yi,Beane:2002vq}. 
A straightforward way to do this is to use interpolating fields that have 
non-zero overlap with the nucleons and deltas. This leads to the tensor
$\c{B}^{ijk}$, which describes a {\bf 70}-dimensional representation of SU$(4|2)$, 
and the tensor $\c{T}^{ijk}$, which describes a {\bf 44}-dimensional representation 
of SU$(4|2)$, see~\cite{Beane:2002vq}.  
The nucleon doublet $N_i$ is embedded in $\c{B}_{ijk}$ as
\begin{equation}
\c{B}_{ijk} = \frac{1}{\sqrt{6}} (\varepsilon_{ij} N_k + \varepsilon_{ik} N_j )
,\end{equation}
where the indices $i$, $j$, and $k$ are restricted to $1$ or $2$. The deltas
are contained simply as $\c{T}^{ijk} = T^{ijk}$ for $i$, $j$, and $k$ restricted
to $1$ or $2$.

To write down the \PQCPT\ Lagrangian in the baryon sector, we must also include
the appropriate grading factors in the contraction of flavor indices. Such factors
are included in the $()$ notation defined in \cite{Labrenz:1996jy,Beane:2002vq}.
The free Lagrangian is given by
\begin{eqnarray}
  \mathfrak{L} &=&
    \left( \ol{\c{B}}\ i\vD\ \c{B} \right)\ 
    +\ 2\am^{\pq} \left( \ol{\c{B}} \c{B} \c{M} \right)\ 
    +\ 2\bm^{\pq} \left( \ol{\c{B}} \c{M} \c{B} \right)\ 
    +\ 2\smb^{\pq} \left( \ol{\c{B}} \c{B} \right) \str(\c{M})
\nonumber\\
 && -
    \left(\ol{\c{T}}^{\mu}\left[\ i\vD - \Delta\ \right] \c{T}_\mu \right)\ 
    +\ 2 \gm^{\pq} \left(\ol{\c{T}}{}^\mu \c{M} \c{T}_\mu \right)\ 
    -\ 2 \smt^{\pq} \left(\ol{\c{T}}{}^\mu \c{T}_\mu \right) \str (\c{M})
\label{eq:leadlagPQ}\,
.\end{eqnarray} 
Above, the covariant derivative's action on either the $\c{B}$ or the $\c{T}$ fields
is defined by
\begin{equation}
(D^\mu \c{B})_{ijk} = \partial^\mu \c{B}_{ijk} + (V^\mu)_{i}{}^{i'} \c{B}_{i'jk} 
+ (-)^{\eta_i(\eta_j + \eta_{j'})} (V^\mu)_{j}{}^{j'} \c{B}_{ij'k} 
+  (-)^{(\eta_i + \eta_j)(\eta_k + \eta_{k'})} (V^\mu)_{k}{}^{k'} \c{B}_{ijk'}
,\end{equation}
and the operators $A^\mu$, $V^\mu$ and $\c{M}$ are all defined analogously 
to those in \CPT. 
The free Lagrangian above contains more operators than the corresponding SU$(2)$ Lagrangian. 
For this reason we have appended $\pq$ superscripts to the coefficients. 
The relations between the free Lagrangian parameters of \PQCPT\
and \CPT\ are found by matching. One finds
\begin{eqnarray}
\a_M &=& \frac{2}{3} \a_M^{\pq} - \frac{1}{3} \b_M^{\pq},  \\
\sigma_M &=& \sigma_M^{\pq} + \frac{1}{6} \a_M^{\pq} + \frac{2}{3} \b_M^{\pq} \\
\gamma_M &=& \gamma_M^{(PQ)} \\
\ol \sigma_M &=& \ol \sigma_M^{(PQ)}
.\end{eqnarray}

The Lagrangian describing the interactions of the {\bf 70} and {\bf 44}
baryons with the pseudo-Goldstone particles is 
\begin{eqnarray}
\mathfrak{L}  &=& 
    2\a\left(\ol{\c{B}} S^\mu \c{B} A_\mu \right)\ 
    +\ 2\beta \left(\ol{\c{B}} S^\mu A_\mu \c{B} \right)\ 
    +\ 2{\cal H} \left( \ol{\c{T}} {}^\nu S^\mu A_\mu \c{T}_\nu \right)
\notag \\
    &+& \sqrt{\frac{3}{2}}{\c{C}} \left[ \left( \ol{\c{T}}{}^\nu A_\nu \c{B}
      \right)\ 
    +\ \left( \ol{\c{B}} A_\nu \c{T}^\nu \right) \right]. 
\end{eqnarray} 
Matching this Lagrangian onto the SU$(2)$ chiral Lagrangian in Eq.~\eqref{eq:MBT}, we find
$\c{H} = g_{\D \D}$ and $\mathcal{C} = - g_{\D N}$. Furthermore, we follow~\cite{Beane:2002vq} 
and choose to express $\a$ and $\b$ in terms of new parameters $g_A$ and $g_1$, namely
\begin{eqnarray}
\a &=& \frac{4}{3} g_A + \frac{1}{3} g_1,  \\
\b &=& \frac{2}{3} g_1 - \frac{1}{3} g_A,
\end{eqnarray}
so that $g_A$ has the same value as in \CPT.

At higher orders in the chiral expansion, the situation in \PQCPT\ parallels that of \CPT\ in Sec.~\ref{CPT}. 
Recall that at higher orders, the Lagrangian can contain arbitrary functions
of $\D / \L_\chi$. We take this into account by implicitly treating 
the leading-order coefficients as functions of $\D / \L_\chi$ expanded out 
to the required order. To maintain the Lorentz invariance of the theory, we use reparameterization invariance 
to generate the higher dimensional operators with fixed coefficients. 
In \PQCPT\ the fixed coefficient Lagrangian is given by
\begin{eqnarray}
\mathfrak{L} &=& 
- \left( \ol{\c{B}} \frac{D_\perp^2}{2 M_B} \c{B} \right)
+ \a \left[ \left( \ol{\c{B}}  \frac{i \loarrow D \cdot S}{M_B}
       \c{B} \, \vit \cdot A \right) - \left( \ol{\c{B}} \frac{S\cdot i \roarrow D}
       {M_B} \c{B} \, \vit \cdot A \right) \right] 
\notag \\
&&
+ \b \left[ \left( \ol{\c{B}}  \frac{i \loarrow D \cdot S}{M_B}
       \vit \cdot A \, \c{B} \right) - \left( \ol{\c{B}} \, \vit\cdot A \frac{S\cdot i \roarrow D}
       {M_B} \c{B} \right) \right] 
+ \left( \ol{\c{T}}^\mu \frac{D_\perp^2}{2 M_B} \c{T}_\mu \right)
\notag \\
&&  
+ \c{H} \left[ \left( \ol{\c{T}}^\mu \frac{i \loarrow D \cdot S}{M_B}
       \vit \cdot A \, \c{T}_\mu \right) - \left( \ol{\c{T}}^\mu \, \vit\cdot A \frac{S\cdot i \roarrow D}
       {M_B} \c{T}_\mu \right) \right] 
\label{eq:fixedPQ}
.\end{eqnarray}

As far as operators with unfixed coefficients, for the nucleon mass we have the 
following partially quenched operators with two insertions of the axial-vector
pion fields
\begin{eqnarray}
\mathfrak{L} &=& \frac{1}{( 4\pi f)}\Bigg\{
        b^{A\pq}_1\, \ol{\c{B}}{}^{kji} \left( A\cdot A \right)_i^{\hskip0.3em n} 
               \c{B}_{njk}
        + b^{A\pq}_2\, (-)^{(\eta_i+\eta_j)(\eta_k+\eta_n)} \ol{\c{B}}^{kji}
           \left( A\cdot A \right)_k^{\hskip0.3em n} \c{B}_{ijn} \nonumber\\
       && 
        + b^{A\pq}_3\, (-)^{\eta_l(\eta_j+\eta_n)} \ol{\c{B}}^{kji} (A_\mu
             )_i^{\hskip0.3em l} (A^\mu )_j^{\hskip0.3em n} \c{B}_{lnk} 
+ 
b_4^{A (PQ)} \, (-)^{\eta_j \eta_n + 1} \ol{\c{B}}^{kji} 
(A_\mu)_i^{\hskip0.3em n} (A^\mu )_j^{\hskip0.3em l} \c{B}_{lnk} 
\notag \\
&& + 
b^{A\pq}_5\, \ol{\c{B}}{}^{kji} \c{B}_{ijk} {\rm Tr}\left( A\cdot A \right)
+ 
b^{vA \pq}_1\, \ol{\c{B}}{}^{kji} \left( v\cdot\, A\, v\cdot A \right)_i^{\hskip0.3em n} \c{B}_{njk}
\notag \\
&& + 
b^{vA \pq}_2\, (-)^{(\eta_i+\eta_j)(\eta_k+\eta_n)} \ol{\c{B}}^{kji}
\left( v\cdot A\, v\cdot A \right)_k^{\hskip0.3emn} \c{B}_{ijn} 
\notag \\
&& +
b^{vA \pq}_3\, (-)^{\eta_l(\eta_j+\eta_n)} \ol{\c{B}}^{kji} 
(v\cdot A )_i^{\hskip0.3em l} (v\cdot A )_j^{\hskip0.3em n} \c{B}_{lnk} 
\notag \\
&& +
b_4^{vA (PQ)} \, (-)^{\eta_j \eta_n + 1} \ol{\c{B}} {}^{kji} 
(v \cdot A)_i^{\hskip0.3em n} (v \cdot A )_j^{\hskip0.3em l} \c{B}_{lnk} 
+ 
b^{vA \pq}_5\, \ol{\c{B}}{}^{kji} \c{B}_{ijk} {\rm Tr}\left( v\cdot A\, v\cdot A \right) 
\label{eq:NhdoPQ} \Bigg\},
\end{eqnarray}
and the following operators with two insertions of the mass operator
\begin{eqnarray}
\mathfrak{L} &=& \frac{1}{( 4\pi f)}\Bigg\{
b^{M \pq}_1\, \ol{\c{B}}{}^{kji} \left( \c{M} \c{M} \right)_i^{\hskip0.3em n} \c{B}_{njk}
+ 
b^{M \pq}_2 \, (-)^{(\eta_i+\eta_j)(\eta_k+\eta_n)}
\ol{\c{B}}{}^{kji} \left( \c{M} \c{M} \right)_k^{\hskip0.3em n} 
               \c{B}_{ijn} \nonumber\\
&& + 
b^{M \pq}_3 \, (-)^{\eta_l(\eta_j+\eta_n)}
               \ol{\c{B}}{}^{kji} (\c{M})_i^{\hskip0.3em l}
               (\c{M})_j^{\hskip0.3em n} \c{B}_{lnk} 
+
b_4^{M \pq} \, (-)^{\eta_j \eta_n + 1 } 
\ol{\c{B}}^{kji} (\c{M})_i^{\hskip0.3em n} (\c{M})_j^{\hskip0.3em l} \c{B}_{lnk} 
\notag \\
&& + 
b^{M \pq}_5\, \ol{\c{B}}{}^{kji} \c{B}_{ijk}\ {\rm str}\left(\c{M}
               \c{M} \right) 
+
b^{M \pq}_6\, \ol{\c{B}}{}^{kji} (\c{M})_i^{\hskip0.3em n} \c{B}_{njk}\ 
               {\rm str}\left( \c{M} \right) 
\notag \\
&& + 
b^{M \pq}_7\, (-)^{(\eta_i+\eta_j)(\eta_k+\eta_n)}
               \ol{\c{B}}{}^{kji} (\c{M})_k^{\hskip0.3em n} \c{B}_{ijn}\ 
               {\rm str}\left( \c{M} \right)
+ 
b^{M \pq}_8 \, \ol{\c{B}}{}^{kji} \c{B}_{ijk}\ [{\rm str}\left( \c{M} \right)]^2
\bigg\}.
\notag \\
\end{eqnarray}%
Matching the operators with coefficients $b_j^{M\pq}$ above onto the SU$(2)$ Lagrangian in Eq.~\eqref{eq:NHDO}, 
we find
%\begingroup \small{
\begin{eqnarray}\label{eq:bmatch1}
b_1^M &=& -\frac{1}{3} b_1^{M \pq} + \frac{2}{3} b_2^{M \pq} - \frac{1}{3} b_3^{M \pq} + \frac{1}{2} b_4^{M \pq}, \\
b_5^M &=&  \frac{2}{3} b_1^{M \pq} + \frac{1}{6} b_2^{M \pq} - \frac{1}{6} b_3^{M \pq} + \frac{1}{6} b_4^{M \pq} 
+ b_5^{M \pq}, \\
b_6^M &=&  \frac{1}{2} b_3^{M \pq} - \frac{1}{3} b_4^{M \pq} - \frac{1}{3} b_6^{M \pq} + \frac{2}{3} b_7^{M \pq}, \\
b_8^M &=&  \frac{1}{6} b_3^{M \pq} - \frac{1}{6} b_4^{M \pq} + \frac{2}{3} b_6^{M \pq} + \frac{1}{6} b_7^{M \pq} 
+ b_8^{M \pq} 
.\end{eqnarray}
%}\endgroup
Carrying out the matching for the operators with coefficients $b_j^{A \pq}$ and $b_j^{vA \pq}$ leads to 
the relations
%\begingroup \small{
\begin{eqnarray}
	b^{A,vA} &=& \frac{1}{2} b_1^{A,vA \pq} 
		+ \frac{1}{2} b_2^{A,vA \pq} 
		- \frac{1}{3} b_3^{A,vA \pq} 
		+ \frac{5}{12} b_4^{A,vA \pq} 
		+ b_5^{A,vA \pq}.
\label{eq:bmatch2}
\end{eqnarray}
%}\endgroup

Higher-dimensional operators that contribute to the masses of the deltas in \PQCPT\ arise similarly from 
two insertions of the axial-vector pion fields
\begin{eqnarray}
	\mathfrak{L} &=& \frac{1}{4 \pi f} 
		\bigg\{ 
			t_1^{A\pq} \, \ol{\c{T}}^{kji}_\mu (A_\nu A^\nu)_{i}^{i'} \c{T}^\mu_{i'jk} 
			+ t_2^A   (-)^{\eta_{i'} (\eta_j + \eta_{j'})}  \ol{\c{T}}^{kji}_\mu (A_\nu)_{i}^{i'} 							(A^\nu)_{j}^{j'} \c{T}^\mu_{i'j'k} \notag \\
		&& 	+ t_3^{A\pq} \, \left( \ol{\c{T}}_\mu \c{T}^\mu \right) \str ( A_\nu A^\nu )
			+ t_1^{\tilde{A}\pq} \, \ol{\c{T}}^{kji}_\mu (A^\mu A_\nu)_{i}^{i'} \c{T}^\nu_{i'jk} 
			+ t_2^{\tilde{A}} \, \ol{\c{T}}^{kji}_\mu (A^\mu)_{i}{}^{i'} (A_\nu)_{j}^{j'} \c{T}^\nu_{i'j'k} 
\notag \\
		&&	+ t_3^{\tilde{A}\pq} \, \left( \ol{\c{T}}_\mu \c{T}^\nu \right) \str ( A^\mu A_\nu )
			+ t_1^{vA\pq} \, \ol{\c{T}}^{kji}_\mu (v \cdot A \, v \cdot A)_{i}^{i'} \c{T}^\mu_{i'jk} 
			+ t_2^{vA} \, \ol{\c{T}}^{kji}_\mu (v \cdot A)_{i}^{i'} (v \cdot A)_{j}^{j'} \c{T}^\mu_{i'j'k} 
\notag \\
		&&	+ t_3^{vA\pq} \, \left( \ol{\c{T}}_\mu \c{T}^\mu \right) \str ( v \cdot A \, v \cdot A )
		\bigg\}
,\label{eq:hdoPQ}
\end{eqnarray}
and two insertions of the mass operator
\begin{eqnarray}\label{eq:tmatch1}
\mathfrak{L} &=&
\frac{1}{4 \pi f} 
\Bigg\{ 
t_1^M \, \ol{\c{T}} {}^{kji}_\mu (\c{M} \c{M})_{i}{}^{i'} \c{T}^\mu_{i'jk} 
+
t_2^M  (-)^{\eta_{i'} (\eta_j + \eta_{j'})} \ol{\c{T}} {}^{kji}_\mu (\c{M})_{i}{}^{i'} (\c{M})_{j}{}^{j'} \c{T}^\mu_{i'j'k} 
\notag \\
&& + 
t_3^M  \, \left( \ol{\c{T}}_\mu \c{T}^\mu \right) \str (\c{M} \c{M})
+
t_4^M \, \left( \ol{\c{T}}_\mu \c{M} \c{T}^\mu \right) \str (\c{M})
+
t_5^M  \, \left( \ol{\c{T}}_\mu \c{T}^\mu \right) [\str(\c{M}\c{M})]^2
\Bigg\}.
\end{eqnarray}
The coefficients of these operators which do not have the superscript, $\pq$, $t_2^{A}$, $t_2^{\tilde{A}}$,
$t_2^{vA}$, and $t_i^{M}$ have the same numerical
values as those in the SU$(2)$ Lagrangian, Eq.~\eqref{eq:hdo}.  The operators with the $\pq$ superscript are related to the $SU(2)$ LECs by the following relations,
\begin{equation}\label{eq:tmatch2}
	t_3^{A,vA,\tilde{A}} =
		\frac{1}{2}t_1^{A,vA,\tilde{A} \pq} +t_3^{A,vA,\tilde{A} \pq}
\end{equation}

%
%
%
%
%
%
%	PQ Nucleons
%
%
%
%
%
%
\subsection{Nucleons}
The relevant diagrams needed to obtain the nucleon mass to NNLO in 
\PQCPT\ are depicted in \cite{Walker-Loud:2004hf} and include
hairpin contributions from the flavor-diagonal states. 
To leading order in \PQCPT\ the nucleon masses are
\begin{equation}
M^{(1)}_B 
= 
\frac{1}{3} \a_M^{(PQ)} m_B^{\prime \prime} 
+ 
\frac{1}{3} \b_M^{(PQ)} m'_B 
+ 
2 \sigma_M^{(PQ)} \, \str (m_Q)
,\end{equation}
where the tree-level coefficients $m^{\prime}_B$ and $m^{\prime \prime}_B$ 
depend on the valence quark masses and are given in Table~\ref{t:mB}.
The next-to-leading contributions to the nucleon masses take the form
\begin{eqnarray}
M^{(3/2)}_B 
& = &
\frac{1}{8 \pi f^2} 
\left[
\sum_\phi A_\phi^B \, m_\phi^3 
+
\sum_{\phi\phi'} A^B_{\phi\phi'} \mathcal{M}^3(m_\phi, m_{\phi'})
\right]
\notag \\
&& + \frac{2 g_{\D N}^2}{(4 \pi f)^2} 
\left[ 
\sum_\phi B_\phi^B  \, \c{F} (m_\phi,\D,\mu)
+
\sum_{\phi\phi'} B^B_{\phi\phi'} \c{F}(m_\phi, m_{\phi'}, \D, \mu)
\right]
,\label{eq:MBPQNLO}
\end{eqnarray}
where $\c{F}(m,\delta,\mu)$ is given in Eq.~\eqref{eqn:F}.
Additionally we have employed the abbreviations
\begin{align}
\mathcal{M}^n (m_\phi, m_{\phi'})  &=  \c{H}_{\phi\phi'}(m^n_\phi, m^n_{\phi'},m^n_X), \notag \\
\c{F} (m_\phi, m_{\phi'}, \d, \mu) &=  
\c{H}_{\phi \phi'}[\c{F}(m_\phi,\d,\mu),\c{F}(m_{\phi'},\d,\mu), \c{F}(m_X,\d,\mu)]
\label{eq:MPQO}
,\end{align}
for contributions arising from the hairpin diagrams. 
The sums involving these two functions are over pairs of flavor-neutral states in the quark basis, e.g., above $\phi\phi'$ runs over
$\eta_u \eta_u$, $\eta_u \eta_d$, and $\eta_d \eta_d$. In this way there is no double counting.
The remaining sums in Eq.~\eqref{eq:MBPQNLO} are over all loop mesons $\phi$ of mass $m_\phi$.
The coefficients $A^B_\phi$, $A^B_{\phi\phi'}$, $B^B_\phi$ and $B_{\phi\phi'}^B$
appear in Table \ref{t:NPQQCD-AB}. 

Finally the next-to-next-to-leading contributions to the nucleon mass are
{\small
\begin{eqnarray}
  M_B^{(2)} &=& ( Z_B - 1) M_B^{(1)} 
    + \frac{1}{4\pi f} \bigg[ \frac{1}{3} b_1^{M \pq} (m^2)_B +
        \frac{1}{6} b_2^{M \pq} (m^2)_B^\prime + \frac{1}{6} b_3^{M
          \pq} (m m')_B +\frac{1}{6} b_4^{M \pq} (m m')_B^\prime \notag \\
&&  +   b_5^{M \pq} \str (m_Q^2) + \frac{1}{3} b_6^{M \pq} m_B^\prime
        \, \str (m_Q) + \frac{1}{6} b_7^{M \pq} m_B^{\prime\prime} \,
        \str(m_Q) + b_8^{M \pq} [\str (m_Q)]^2 \bigg] \notag \\
&&  - \frac{1}{(4 \pi f)^2} \left[ \sum_{\phi} C^B_\phi \c{L}
        (m_\phi,\mu) 
        + \sum_{\phi \phi'} C^B_{\phi,\phi'} \c{L}(m_\phi,m_{\phi'},\mu)
        \right] \notag \\
&&  - \frac{2 \sigma_M^{(PQ)}}{(4 \pi f)^2} 
       \left[ \sum_\phi \ol C_\phi \c{L} (m_\phi,\mu)
        + \sum_{\phi \phi'} \ol C_{\phi \phi'} \c{L} (m_\phi,
         m_{\phi'},\mu) \right] \notag \\
&&  + \frac{1}{(4 \pi f)^3} \sum_{\phi} \left( 
        b_1^{A \pq} D^B_{\b,\phi} + b_2^{A \pq} D^B_{\a,\phi} + b_3^{A
          \pq} E_\phi^B + b_4^{A \pq} E^{\prime B}_\phi + b_5^{A \pq}
          \ol D_\phi \right) 
        \ol{\c{L}} (m_\phi, \mu) \notag \\
&&  + \frac{1}{(4 \pi f)^3} \sum_{\phi \phi'} \left( 
        b_1^{A \pq} D^B_{\b,\phi\phi'} + b_2^{A \pq}
         D^B_{\a,\phi\phi'} + b_3^{A \pq} E_{\phi\phi'}^B + b_4^{A \pq}
         E^{\prime B}_{\phi \phi'} 
        + b_5^{A \pq} \ol D_{\phi\phi'} \right) 
        \ol{\c{L}} (m_\phi, m_{\phi'}, \mu) \notag \\
&&  + \frac{1}{4 (4 \pi f)^3} \sum_\phi  \left( 
        b_1^{vA \pq} D^B_{\b,\phi} + b_2^{vA \pq} D^B_{\a,\phi} 
        + b_3^{vA \pq} E_\phi^B + b_4^{vA \pq} E_{\phi}^{\prime B}  
        + b_5^{vA \pq} \ol D_\phi \right) \notag \\
&& \phantom{bigspacethattakesupspace}
        \times \left[ \ol{\c{L}} (m_\phi, \mu) - \frac{1}{2} m_\phi^4
        \right] \notag \\
&&  + \frac{1}{4 (4 \pi f)^3} \sum_{\phi \phi'} \left( 
        b_1^{vA \pq} D^B_{\b,\phi\phi'} + b_2^{vA \pq}
         D^B_{\a,\phi\phi'} + b_3^{vA \pq} E_{\phi\phi'}^B 
        + b_4^{vA \pq} E_{\phi \phi'}^{\prime B} 
        + b_5^{vA \pq} \ol D_{\phi\phi'} \right) \notag \\
&& \phantom{bigspacethattakesupspace}
        \times \left[ \ol{\c{L}} \left( m_\phi, m_{\phi'}, \mu \right)
          - \frac{1}{2} \mathcal{M}^4(m_\phi, m_{\phi'}) \right] \notag \\
&&  + \frac{9}{8 M_B} \frac{1}{(4 \pi f)^2} \left\{ 
        \sum_\phi A^B_\phi \left[ \ol{\c{L}} (m_\phi, \mu) + \frac{5}{6}
          m_\phi^4 \right]
        + \sum_{\phi \phi'} A^B_{\phi \phi'} \left[
          \ol{\c{L}} \left( m_\phi, m_{\phi'}, \mu \right)
          + \frac{5}{6} \mathcal{M}^4(m_\phi, m_{\phi'})
        \right]\right\} \notag \\
&&  +  \frac{15}{8 M_B} \frac{g_{\D N}^2}{(4 \pi f)^2} \left\{
         \sum_\phi B^B_\phi \left[ \ol{\c{L}} (m_\phi, \mu) +
           \frac{9}{10} m_\phi^4 \right] 
         +\ \sum_{\phi \phi'} B^B_{\phi \phi'} \left[
           \ol{\c{L}} \left( m_\phi, m_{\phi'}, \mu \right)
           + \frac{9}{10} \mathcal{M}^4(m_\phi, m_{\phi'})
         \right]\right\} \notag \\
&& + 
\frac{ 6 \sigma_M^{(PQ)} \str (m_Q)}{(4 \pi f)^2} 
\left\{
\sum_\phi A^B_\phi 
\left[ \c{L} (m_\phi, \mu) + \frac{2}{3} m_\phi^2 \right]
+ 
\sum_{\phi \phi'} A^B_{\phi \phi '} 
\left[
\c{L} \left( m_\phi, m_{\phi'}, \mu \right)
+ \frac{2}{3} \mathcal{M}^2(m_\phi,m_{\phi'})
\right]
\right\}
\notag \\
&& +
\frac{6 g_{\D N}^2 \ol \sigma_M \str(m_Q)}{(4 \pi f)^2} 
\left\{ 
\sum_\phi B^B_\phi 
\left[ \c{J} (m_\phi, \D, \mu) + m_\phi^2 \right]
+
\sum_{\phi \phi'} B^B_{\phi \phi'} 
\left[
\c{J} (m_\phi, m_{\phi'}, \D, \mu)
+
\mathcal{M}^2(m_\phi, m_{\phi'})
\right]
\right\}
\notag \\
&& + 
\frac{3}{ (4 \pi f)^2} 
\left\{
\sum_\phi F^B_\phi 
\left[ \c{L} (m_\phi, \mu) + \frac{2}{3} m_\phi^2 \right]
+
\sum_{\phi \phi'} F^B_{\phi \phi'}
\left[
\c{L} \left( m_\phi, m_{\phi'}, \mu \right)
+ 
\frac{2}{3} \mathcal{M}^2(m_\phi,m_{\phi'})
\right]
\right\}
\notag \\
&& - 
\frac{2 g_{\D N}^2 \g_M}{(4 \pi f)^2} 
\left\{
\sum_\phi G^B_\phi 
\left[ \c{J} (m_\phi, \D, \mu) + m_\phi^2 \right]
+
\sum_{\phi \phi'} G^B_{\phi \phi'}
\left[
\c{J}(m_\phi, m_{\phi'}, \D, \mu)
+
\mathcal{M}^2(m_\phi, m_{\phi'})
\right]
\right\}
\label{eq:NMPQNNLO}
\end{eqnarray}
}

The above expression is written quite compactly. It involves the 
tree-level coefficients $m_B^\prime$, $m_B^{\prime \prime}$, $(m^2)_B$, $(m^2)_B^\prime$, $(mm')_B$, 
and $(mm')_B^\prime$. These are listed in Table~\ref{t:mB} for the nucleon states labeled by $B$.  
The wavefunction renormalization $Z_B$ appearing in the above expression
is given by
{\small
\begin{eqnarray}
Z_B - 1 
&=& 
   - \frac{3}{(4 \pi f )^2}
\left\{ 
\sum_\phi A^B_\phi 
\left[  \c{L}(m_\phi , \mu) + \frac{2}{3} m_\phi^2 \right]
+
\sum_{\phi \phi'} A^B_{\phi \phi'}
\left[
\c{L} \left( m_\phi, m_{\phi'},\mu  \right)
+
\frac{2}{3} \mathcal{M}^2(m_\phi, m_{\phi'})
\right]
\right\}
\notag \\
&&- 
\frac{3 g_{\D N}^2}{(4 \pi f)^2 } 
\left\{ 
\sum_\phi B^B_\phi \left[ \c{J}(m_\phi, \D, \mu) + m_\phi^2 \right]
+
\sum_{\phi \phi'} B^B_{\phi \phi'} 
\left[
\c{J}(m_\phi,m_{\phi'},\D,\mu)
+
\mathcal{M}^2(m_\phi, m_{\phi'})
\right]
\right\} 
\notag \\
\label{eq:ZBPQ} \end{eqnarray}
}
Furthermore we have made use of abbreviations for the new non-analytic 
functions arising from loop integrals in Eqs.~\eqref{eq:NMPQNNLO} and
\eqref{eq:ZBPQ}. These are defined to be
\begin{align}
\c{L}(m_\phi,m_{\phi'},\mu) &= \c{H}_{\phi\phi'}
[ \c{L}(m_\phi,\mu),\c{L}(m_{\phi'}, \mu), \c{L}(m_X,\mu)], \notag \\   
\ol{\c{L}}(m_\phi,m_{\phi'},\mu) &= \c{H}_{\phi\phi'}
[ \ol{\c{L}}(m_\phi,\mu), \ol{\c{L}}(m_{\phi'}, \mu), \ol{\c{L}}(m_X,\mu)], \notag \\ 
\c{J}(m_\phi,m_{\phi'},\d,\mu) &= \c{H}_{\phi\phi'} 
[\c{J}(m_\phi,\d,\mu),\c{J}(m_{\phi'},\d,\mu),\c{J}(m_X,\d,\mu) ]
\label{hairpin} \end{align}
and arise from hairpin contributions. 
The various coefficients in the above sums over loop mesons and loop pairs of flavor-neutral mesons 
are listed in Tables \ref{t:NPQQCD-AB}--\ref{t:NPQQCD-G}. 
One may check that in the limit $m_j \to m_u$, $m_l \to m_d$ the \CPT\ results are obtained from 
the \PQCPT\ expressions with the proper matching of coefficients, see
Eq.~\eqref{eq:bmatch1}-\eqref{eq:tmatch2}.

%
%
%
%
%
%
%	PQ Deltas
%
%
%
%
%
%
\subsection{Deltas}

The relevant diagrams needed to obtain the delta masses at NNLO in \PQCPT\
are depicted in~\cite{Tiburzi:2004rh}. These include hairpin contributions
from the flavor-diagonal propagator. 
To leading order in \PQCPT\ the delta masses are
\begin{equation}
M^{(1)}_T = \frac{2}{3} \gamma_M \, m_T  - 2 \, \ol \sigma_M \, \str (m_Q),
\label{eq:MPQLO}
\end{equation}
where the coefficients  $m_T$ are listed for deltas $T$ in Table~\ref{t:mT}. 
At next-to-leading order, we have
\begin{eqnarray}
M^{(3/2)}_T &=& -\frac{5 g_{\D \D}^2}{72 \pi f^2} 
\left[ 
\sum_\phi A_\phi^T \, m_\phi^3 
+ 
\sum_{\phi\phi'} A_{\phi\phi'}^T \, \mathcal{M}^3 (m_\phi, m_{\phi'}) 
\right]
\notag \\ 
&-& \frac{g_{\D N}^2}{(4 \pi f)^2} 
\left[ \sum_\phi  B_\phi^T  \, \c{F} (m_\phi,-\D,\mu)
+
\sum_{\phi\phi'} B^T_{\phi\phi'} \, \c{F} (m_\phi, m_{\phi'}, -\D, \mu) \right]
,\end{eqnarray}
where $\mathcal{F}(m,\D,\mu)$ is given by Eq.~\eqref{eqn:F}
and the functions $\mathcal{M}^n (m_\phi, m_{\phi'})$
and $\c{F} (m_\phi, m_{\phi'}, \d, \mu)$ appear in 
Eq.~\eqref{eq:MPQO}.
The coefficients $A^T_\phi$, $A^T_{\phi\phi'}$, $B^T_\phi$ and
$B_{\phi\phi'}^T$ appear in Table \ref{t:PQQCD-A}.

At next-to-next-to leading order, we have the following contribution to the delta masses
{\small
\begin{eqnarray}
	M_T^{(2)} &=& (Z_T - 1) M_T^{(1)} \notag\\
	&&	+ \frac{1}{4\pi f} \bigg[ \frac{t_1^M}{3} (m^2)_T 
			+ \frac{t_2^M}{3} (m m')_T 
			+ t_3^M \, \str(m_Q^2) 
			+ \frac{ t_4^M}{3}   m_T  \, \str (m_Q)
			+ t_5^M \, [\str (m_Q)]^2 \bigg] \notag \\
	&&	-\frac{2 \,\g_M}{(4 \pi f)^2} \left[
			\sum_\phi C_\phi^T \, \c{L}(m_\phi,\mu) 
			+ \sum_{\phi\phi'} C_{\phi\phi'}^T \, \c{L}(m_\phi,m_{\phi'},\mu) \right] \notag \\
	&& 	+ \frac{2 \, \ol{\s}_M}{(4 \pi f)^2} \left[
			\sum_\phi \ol C_\phi \, \c{L}(m_\phi,\mu)
			+ \sum_{\phi\phi'} \ol C_{\phi\phi'} \, \c{L}(m_\phi,m_{\phi'},\mu) \right] \notag \\
	&& 	+ \frac{1}{(4 \pi f )^3} \sum_\phi \left( 
			t_1^{A\pq} \,  D_\phi^T 
			+ t^A_2 \, E_\phi^T 
			+ t_3^{A\pq} \, \ol D_\phi \right) \ol{\c{L}} (m_\phi, \mu) \notag \\
	&& 	+ \frac{1}{(4 \pi f )^3} \sum_{\phi\phi'} \left( 
			t_1^{A\pq} \,  D_{\phi\phi'}^T 
			+ t^A_2 \, E_{\phi\phi'}^T 
			+ t_3^{A\pq} \, \ol D_{\phi\phi'} \right) \ol{\c{L}} (m_\phi,m_{\phi'}, \mu) \notag \\
	&& 	+ \frac{1}{4 (4 \pi f )^3} \sum_\phi \left[ 
			(t^{\tilde{A}\pq}_1 
			+ t_1^{vA\pq} ) D_\phi^T 
			+  (t_2^{\tilde{A}} 
			+ t_2^{vA} ) E_\phi^T 
			+ (t_3^{\tilde{A}\pq} 
			+ t_3^{vA\pq} ) \ol D_\phi \right] \notag\\
	&& \qquad\qquad\qquad \times
		\left[ \ol{\c{L}} (m_\phi, \mu) -\frac{1}{2} m_\phi^4 \right] \notag \\
	&& 	+ \frac{1}{4 (4 \pi f )^3} \sum_{\phi\phi'} \left[ 
			(t_1^{\tilde{A\pq}} 
			+ t_1^{vA\pq} ) D_{\phi\phi'}^T 
			+ (t^{\tilde{A}\pq}_2 
			+ t_2^{vA} ) E_{\phi\phi'}^T 
			+ (t_3^{\tilde{A}\pq} 
			+ t_3^{vA}) \ol{D}_{\phi\phi'} \right] \notag\\
	&& \qquad\qquad\qquad \times
			\left[ \ol{\c{L}} \left(m_\phi,m_{\phi'}, \mu \right)
				- \frac{1}{2} \mathcal{M}^4(m_\phi, m_{\phi'}) \right] \notag \\
	&&	 - \frac{5}{8} \frac{g_{\D \D}^2}{(4\pi f)^2 M_B} \left\{ 
			\sum_\phi A_\phi^T 
\left[ 
\ol{\c{L}}(m_\phi,\mu) + \frac{19}{10} m_\phi^4
\right]
+
\sum_{\phi\phi'} A_{\phi\phi'}^T 
\left[
\ol{\c{L}} \left( m_\phi, m_{\phi'}, \mu \right)
+ 
\frac{19}{10} \mathcal{M}^4(m_\phi, m_{\phi'})
\right]
\right\}
\notag \\
&& - 
\frac{15}{16} \frac{g_{\D N}^2}{(4 \pi f)^2 M_B}  
\left\{
\sum_\phi B_\phi^T 
\left[ 
\ol{\c{L}}(m_\phi,\mu)
- \frac{1}{10} m_\phi^4
\right]
+ 
\sum_{\phi\phi'} B_{\phi\phi'}^T
\left[
\ol{\c{L}}\left( m_\phi, m_{\phi'}, \mu \right)
-
\frac{1}{10} \mathcal{M}^4(m_\phi, m_{\phi'})
\right]
\right\}
\notag \\
&& -
\frac{10 g_{\D \D}^2 \ol \sigma_M \str(m_Q)}{3 (4 \pi f)^2} 
\left\{
\sum_\phi A_\phi^T 
\left[ 
\c{L}(m_\phi,\mu) + \frac{26}{15} m_\phi^2
\right]
+ 
\sum_{\phi \phi'} A_{\phi\phi'}^T 
\left[
\c{L} \left( m_\phi, m_{\phi'}, \mu \right)
+ 
\frac{26}{15} \mathcal{M}^2 (m_\phi, m_{\phi'})
\right]
\right\}
\notag \\
&&- 
\frac{3 g_{\D N}^2 \sigma_M^{(PQ)} \str(m_Q) }{(4 \pi f)^2} 
\left[
\sum_\phi B_\phi^T  \, \c{J} (m_\phi,-\D,\mu)
+
\sum_{\phi\phi'} B_{\phi\phi'}^T \, \c{J}(m_\phi,m_{\phi'},-\D,\mu)
\right]
\notag \\
&& + \frac{10 g_{\D \D}^2 \g_M}{9 (4 \pi f)^2}
	  \left\{ \sum_\phi F_\phi^T
	   \left[ \c{L}(m_\phi,\mu) + \frac{26}{15} m_\phi^2  \right]
	 + \sum_{\phi \phi'} F_{\phi\phi'}^T 
	     \left[ \c{L} \left(m_\phi, m_{\phi'}, \mu \right) 
	     +  \frac{26}{15} \mathcal{M}^2(m_\phi, m_{\phi'}) \right] \right\} \notag \\
&&- 
\frac{3 g_{\D N}^2 }{2 (4 \pi f)^2} 
\left[
\sum_\phi G_\phi^T  \, \c{J} (m_\phi,-\D,\mu)
+
\sum_{\phi\phi'} G_{\phi\phi'}^T \, \c{J}(m_\phi,m_{\phi'},-\D,\mu)
\right]
\label{eq:MPQNNLO}
.\end{eqnarray}}
As with the NNLO contributions to the nucleon masses, the above expression is written quite compactly. 
It involves the tree-level coefficients $m_T$, $(m^2)_T$, and $(mm')_T$
which are listed in Table~\ref{t:mT} for the delta states $T$. 
The NNLO result involves the wavefunction renormalization $Z_T$, which 
we find to be
{\small
\begin{eqnarray}
Z_T - 1 &=& - \frac{5 g_{\D \D}^2}{3 (4 \pi f)^2} 
\left\{
\sum_\phi A_\phi^T 
\left[ \c{L}(m_\phi,\mu) + \frac{26}{15} m_\phi^2 \right]
+ 
\sum_{\phi\phi'} A_{\phi\phi'}^T 
\left[
\c{L} \left( m_\phi,m_{\phi'},\mu \right) 
+
\frac{26}{15} \mathcal{M}^2(m_\phi, m_{\phi'})
\right]
\right\}
\notag \\
&&- \frac{3 g_{\D N}^2}{2 (4 \pi f)^2} 
\left[
\sum_\phi B_\phi^T  \c{J} (m_\phi,-\D,\mu)
+
\sum_{\phi\phi'} B_{\phi\phi'}^T \c{J}(m_\phi,m_{\phi'},-\D,\mu)
\right].
\label{eq:PQwfn}
\end{eqnarray}}
The functions arising from hairpin contributions are listed in Eq.~\eqref{hairpin}. 
The various coefficients in the above sums are listed in Tables \ref{t:PQQCD-A}--\ref{t:PQQCD-G}. 
Finally, one can check that the \CPT\ result is recovered in the limit $m_j \to m_u$, $m_l \to m_d$, and by utilizing the matching relations given in Eqs.~\eqref{eq:bmatch1},~\eqref{eq:bmatch2},~\eqref{eq:tmatch1} and \eqref{eq:tmatch2}.

\bibliography{MasterBib}

%
%
%
%
%
%
%	PQ Tables
%
%
%
%
%
%
\section{Partially quenched coefficient tables}\label{ap:N}

Here we list the tables needed for the partially quenched calculation
which have not already been listed in the text.  One can obtain from the authors a Mathematica notebook which evaluates \PQCPT\ and \CPT\ 
expressions for the nucleon and delta masses in various limits.  We
consider the case of non-degenerate valence and sea quarks, as well as
non-degenerate valence and degenerate sea quarks and further
degenerate valence and sea quarks.

\begin{table}
\caption{The tree-level coefficients in \CPT\ and \PQCPT. The
  coefficients $m_B$, $m_B^\prime$, $m_B^{\prime \prime}$, $(m^2)_B$,
  $(m^2)_B^\prime$, $(mm')_B$, and $(mm')_B^\prime$ are listed for the
  nucleons $B$.}
%\begin{ruledtabular}
\begin{tabular}{l | c c c c c c c}
 & $\quad m_B \quad $ & $\qquad m_B^\prime \quad$ & $\qquad m_B^{\prime \prime} \quad$ 
 & $ \qquad (m^2)_B \quad $ & $\qquad (m^2)_B^\prime \quad $ & $\qquad (m m' )_B \quad$ 
 & $\qquad (mm')_B^\prime \quad$  \\
\hline
$p$
 & $m_u$ 
 & $m_u + 2 m_d$ 
 & $5 m_u + m_d$ 
 & $m_u^2 + 2 m_d^2$ 
 & $5 m_u^2 + m_d^2$ 
 & $m_u^2 + 5 m_u m_d$ 
 & $  m_u^2 - 4 m_u m_d$ \\    
$n$
 & $m_d$ 
 & $2 m_u + m_d$ 
 & $m_u + 5 m_d$ 
 & $2 m_u^2 + m_d^2$ 
 & $m_u^2 + 5 m_d^2$ 
 & $5 m_u m_d + m_d^2$ 
 & $- 4 m_u m_d + m_d^2$
\end{tabular}
%\end{ruledtabular}
\label{t:mB}
\end{table}

%\begingroup
%\squeezetable
\begin{table}[hb]
\caption{The coefficients $A^B_\phi$, $A^B_{\phi\phi'}$, $B^B_\phi$, and $B^T_{\phi\phi'}$ in \PQCPT. 
Coefficients are listed for the nucleons $B$, and for $A^B_\phi$ and $B^B_\phi$
are grouped into contributions from loop mesons
with mass $m_\phi$, while for $A^T_{\phi\phi'}$ and $B^B_{\phi\phi'}$ 
are grouped into contributions from pairs of quark-basis $\eta_q$ mesons.}
%\begin{ruledtabular}
\begin{tabular}{l | c c c c }
 & \multicolumn{4}{c}{$A^B_\phi \phantom{ap}$} \\
 & $\quad \eta_u \quad$ 
 & $\quad \pi^\pm \quad $ 
 & $\quad  \eta_d \quad $ \\
\hline

$p$
  & $\frac{1}{3}( g_A^2 + 2 g_A g_1+ g_1^2/4)$
  & $\frac{1}{3} (g_A^2 - g_A g_1- 5 g_1^2 / 4)$
  & $0$\\
$n$
  & $0$  
  & $\frac{1}{3} (g_A^2 - g_A g_1- 5 g_1^2 / 4)$  
  &  $\frac{1}{3}( g_A^2 + 2 g_A g_1+ g_1^2/4)$\\

\multicolumn{4}{c}{}\\
 & $ \quad ju \quad$ 
 & $ \quad lu \quad$ 
 & $\quad jd \quad$  
 & $\quad ld \quad$ \\
\hline

$p$
  & $\frac{1}{3}(2 g_A^2 + g_A g_1+ g_1^2/2)$ 
  & $\frac{1}{3}(2 g_A^2 + g_A g_1+ g_1^2/2)$  
  &  $g_1^2 /4 $ 
  & $g_1^2 /4$ \\
$n$
  &  $g_1^2 /4$ 
  & $g_1^2 /4$ 
  &  $\frac{1}{3}(2 g_A^2 + g_A g_1+ g_1^2/2)$ 
  & $\frac{1}{3}(2 g_A^2 + g_A g_1+ g_1^2/2)$  \\

\multicolumn{4}{c}{}\\
&\multicolumn{4}{c}{$A^B_{\phi\phi'}$} \\
  & $\quad \eta_u \eta_u \quad $ 
  & $\quad \eta_u \eta_d\quad $   
  & $\quad \eta_d \eta_d\quad$ \\
\hline
$p$
  & $g_A^2 + g_A g_1 + g_1^2/4$ 
  & $g_A g_1 + g_1^2/2$ 
  & $g_1^2/4$ \\
$n$
  & $g_1^2/4$  
  & $g_A g_1 + g_1^2/2$ 
  & $g_A^2 + g_A g_1 + g_1^2/4$ \\

\multicolumn{4}{c}{}\\
\multicolumn{4}{c}{}\\

& \multicolumn{4}{c}{$B^B_\phi \phantom{ap}$} \\ 
  & $\quad \eta_u \quad$ 
  & $\quad \pi^\pm \quad $ 
  & $\quad  \eta_d \quad $\\
\hline

$p$
  & $\frac{1}{9}$ 
  & $\frac{5}{9}$  
  & $0$\\ 
$n$
  & $0$ 
  & $\frac{5}{9}$  
  & $\frac{1}{9}$\\

\multicolumn{4}{c}{}\\
  & $ \quad ju \quad$ 
  & $ \quad lu \quad$ 
  & $\quad jd \quad$  
  & $\quad ld \quad$ \\
\hline
$p$
    & $\frac{1}{9}$ 
  & $\frac{1}{9}$  
  & $\frac{2}{9}$ 
  & $\frac{2}{9}$ \\
$n$
  & $\frac{2}{9}$ 
  & $\frac{2}{9}$  
  & $\frac{1}{9}$ 
  & $\frac{1}{9}$ \\

\multicolumn{4}{c}{}\\
  &\multicolumn{4}{c}{$B^B_{\phi\phi'}$ \phantom{sp}} \\
  & $\quad \eta_u \eta_u \quad $ 
  & $\quad \eta_u \eta_d\quad $   
  & $\quad \eta_d \eta_d\quad$ \\
\hline
$p$
  & $\frac{2}{9}$ 
  & $-\frac{4}{9}$ 
  & $\frac{2}{9}$ \\
$n$
  & $\frac{2}{9}$ 
  & $-\frac{4}{9}$ 
  & $\frac{2}{9}$ \\
\end{tabular}
%\end{ruledtabular}
\label{t:NPQQCD-AB}
\end{table}
%\endgroup

\begingroup
\squeezetable
%\begin{turnpage}
\begin{table}[ht]
\caption{The coefficients $C^B_\phi$, $C^B_{\phi\phi'}$, $\ol C_\phi$,
  $\ol C_{\phi \phi'}$, $D^B_{\a,\phi}$, $D^B_{\a,\phi\phi'}$,
  $D^B_{\b,\phi}$, $D^B_{\b,\phi\phi'}$, $\ol D_\phi$, and $\ol
  D_{\phi \phi'}$ in \PQCPT. The coefficients $\ol C_\phi$,
  $\ol C_{\phi \phi'}$, $\ol D_\phi$ and $\ol
  D_{\phi \phi'}$ are identical for all nucleons and
  deltas. The remaining coefficients are listed for nucleon states
  $B$, and for $C^B_\phi$, and $D^B_\phi$ are grouped into
  contributions from loop  mesons with mass $m_\phi$, while for
  $C^B_{\phi\phi'}$ and $D^B_{\phi\phi'}$ are grouped into
  contributions from pairs of quark-basis $\eta_q$ mesons. If a
  particular meson or pair of flavor-neutral  mesons is not listed,
  then the value of the coefficient is zero for the nucleons. }
%\begin{ruledtabular}
\begin{tabular}{l | c c c c }
& \multicolumn{4}{c}{$C^B_\phi $}\\
  & $\quad ju \quad$ 
  & $\quad lu \quad $ 
  & $\quad jd \quad $  
  & $\quad ld \quad$\\
\hline

$p$
  & $\frac{1}{3} (5 \a_M^{\pq} + 2 \b_M^{\pq})$ 
  & $\frac{1}{3} (5 \a_M^{\pq} + 2 \b_M^{\pq})$  
  & $\frac{1}{3} (\a_M^{\pq} + 4 \b_M^{\pq})$ 
  & $\frac{1}{3} (\a_M^{\pq} + 4 \b_M^{\pq})$\\

  & $\quad \times (m_u+ m_j)$ 
  & $\quad \times (m_u + m_l)$ 
  & $\quad \times (m_d + m_j)$ 
  & $\quad \times (m_d + m_l)$\\
\\
$n$ 
  & $\frac{1}{3} (\a_M^{\pq} + 4 \b_M^{\pq})$ 
  & $\frac{1}{3} (\a_M^{\pq} + 4 \b_M^{\pq})$  
  & $\frac{1}{3} (5 \a_M^{\pq} + 2 \b_M^{\pq})$ 
  & $\frac{1}{3} (5 \a_M^{\pq} + 2 \b_M^{\pq})$\\

  & $\quad \times (m_u+ m_j)$ 
  & $\quad \times (m_u + m_l)$ 
  & $\quad \times (m_d + m_j)$ 
  & $\quad \times (m_d + m_l)$\\
\end{tabular}

\bigskip

\begin{tabular}{l | c c }
&\multicolumn{2}{c}{$C^B_{\phi\phi'}$} \\
  & $\quad \eta_u \eta_u \quad$ 
  & $\quad \eta_d \eta_d \quad$ \\
\hline
$p$
  & $\frac{2}{3} (5 \a_M^{\pq} + 2 \b_M^{\pq}) m_u$ 
  & $\frac{2}{3} (\a_M^{\pq} + 4 \b_M^{\pq}) m_d$ \\ \\
$n$
  & $\frac{2}{3} (\a_M^{\pq} + 4 \b_M^{\pq}) m_u$ 
  & $\frac{2}{3} (5 \a_M^{\pq} + 2 \b_M^{\pq}) m_d$ \\
\end{tabular}

\bigskip

\begin{tabular}{ c c c | c c}
    \multicolumn{3}{c|}{$\ol C_\phi$}
  & \multicolumn{2}{c}{$\ol C_{\phi\phi'}$}\\
    $\quad jj \quad$ 
  & $\quad jl \quad$ 
  & $\quad ll \quad$
  & $\quad \eta_j \eta_j \quad$ 
  & $\quad \eta_l \eta_l \quad$ \\
\hline
    $2 m_j$
  & $2 (m_j +m_l)$
  & $2 m_l$
  & $2 m_j$ 
  & $2 m_l$\\
\end{tabular}

\bigskip

\begin{tabular}{l | c c c c | c c}
& \multicolumn{4}{c|}{$D^B_{\a,\phi}$} 
& \multicolumn{2}{c}{$D^B_{\a,\phi\phi'}$} \\
  & $\quad ju \quad$ 
  & $\quad lu \quad $ 
  & $\quad jd \quad $  
  & $\quad ld \quad$ 
  & $\quad \eta_u \eta_u \quad$ 
  & $\quad \eta_d \eta_d \quad$ \\
\hline
$p$       
  & $\frac{5}{6}$ 
  & $\frac{5}{6}$  
  & $\frac{1}{6}$ 
  & $\frac{1}{6}$ 
  & $\frac{5}{6}$ 
  & $\frac{1}{6}$ \\

$n$ 
  & $\frac{1}{6}$ 
  & $\frac{1}{6}$  
  & $\frac{5}{6}$ 
  & $\frac{5}{6}$
  & $\frac{1}{6}$ 
  & $\frac{5}{6}$ \\

\multicolumn{6}{c}{}
\\
& \multicolumn{4}{c|}{$D^B_{\b,\phi} $} 
& \multicolumn{2}{c}{$D^B_{\b,\phi\phi'}$} \\
  & $\quad ju \quad$ 
  & $\quad lu \quad $ 
  & $\quad jd \quad $  
  & $\quad ld \quad$ 
  & $\quad \eta_u \eta_u \quad$ 
  & $\quad \eta_d \eta_d \quad$ \\
\hline
$p$       
  & $\frac{1}{3}$ 
  & $\frac{1}{3}$  
  & $\frac{2}{3}$ 
  & $\frac{2}{3}$ 
  & $\frac{1}{3}$ 
  & $\frac{2}{3}$ \\

$n$ 
  & $\frac{2}{3}$ 
  & $\frac{2}{3}$  
  & $\frac{1}{3}$ 
  & $\frac{1}{3}$
  & $\frac{2}{3}$ 
  & $\frac{1}{3}$ \\
\end{tabular}

\bigskip

\begin{tabular}{c c c | c c}
    \multicolumn{3}{c|}{$\ol D_\phi$}
  & \multicolumn{2}{c}{$\ol D_{\phi\phi'}$}\\
    $\quad jj \quad$ 
  & $\quad jl \quad$ 
  & $\quad ll \quad$ 
  & $\quad \eta_j \eta_j \quad$ 
  & $\quad \eta_l \eta_l \quad$ \\  
\hline 
    $1$ 
  & $2$ 
  & $1$ 
  & $1$ 
  & $1$
\end{tabular}
%\end{ruledtabular}
\label{t:NPQQCD-C}
\end{table}
%\end{turnpage}

\begin{table}[ht]
\caption{The coefficients $E^B_\phi$, $E^B_{\phi\phi'}$, $E^{\prime
    B}_\phi$ and $E^{\prime B}_{\phi \phi'}$ in \PQCPT. Coefficients
  are listed for the nucleons, and for $E^B_\phi$ and $E^{\prime
    B}_\phi$ are grouped into contributions from loop mesons with mass
  $m_\phi$, while for $E^B_{\phi\phi'}$ and $E^{\prime B}_{\phi
    \phi'}$ are grouped into contributions from pairs of quark-basis
  $\eta_q$ mesons.}
%\begin{ruledtabular}
\begin{tabular}{l | c c c | c  c c}
& \multicolumn{3}{c|}{$E^B_\phi \phantom{ap}$} & \multicolumn{3}{c}{$E^B_{\phi\phi'}$ \phantom{sp}} \\
& $\quad \eta_u \quad$ & $\quad \pi^\pm \quad $ & $\quad \eta_d \quad $  
& $\quad \eta_u \eta_u \quad$ & $\quad \eta_u \eta_d \quad $& $\quad \eta_d \eta_d \quad$ \\
\hline
$p$       
           &  $\frac{1}{6}$ & $-\frac{2}{3}$  & $0$ 
           &  $\frac{1}{6}$ & $ \frac{5}{6}$  & $0$ \\

$n$ 
           &  $0$ & $-\frac{2}{3}$  & $\frac{1}{6}$ 
           &  $0$ & $ \frac{5}{6}$  & $\frac{1}{6}$ \\

\multicolumn{6}{c}{}
\\
& \multicolumn{3}{c|}{$E^{\prime B}_\phi \phantom{ap}$} & \multicolumn{3}{c}{$E^{\prime B}_{\phi\phi'}$ \phantom{sp}} \\
& $\quad \eta_u \quad$ & $\quad \pi^\pm \quad $ & $\quad \eta_d \quad $  
& $\quad \eta_u \eta_u \quad$ & $\quad \eta_u \eta_d \quad $& $\quad \eta_d \eta_d \quad$ \\
\hline
$p$       
           &  $\frac{1}{6}$ & $ \frac{5}{6}$  & $0$ 
           &  $\frac{1}{6}$ & $-\frac{2}{3}$  & $0$ \\

$n$ 
           &  $0$ & $ \frac{5}{6}$  & $\frac{1}{6}$ 
           &  $0$ & $-\frac{2}{3}$  & $\frac{1}{6}$ \\
\end{tabular}
%\end{ruledtabular}
\label{t:NPQQCD-E}
\end{table}
\endgroup

%\begin{turnpage}
\begingroup
\squeezetable
\begin{table}[ht]
\caption{The coefficients $F^B_\phi$ and $F^B_{\phi\phi'}$ in \PQCPT. Coefficients are
listed for the nucleons, and for $F^B_\phi$ are grouped into contributions from loop mesons
with mass $m_\phi$, while for $F^B_{\phi\phi'}$ are grouped into contributions from pairs of quark-basis 
$\eta_q$ mesons.}
%\begin{ruledtabular}
\begin{tabular}{l | l l l l  }
& \multicolumn{4}{c}{$F^B_\phi \phantom{ap}$} \\
      & $\qquad \qquad \eta_u $ & $\qquad \qquad \pi^\pm $ & $\qquad \quad \eta_d $ & \\  
\hline
\hline

$p$       
&  $\frac{1}{24} \{ m_u [16 g_A^2 \a_M^{\pq} $ 
&  $\frac{1}{24} \{ m_u [16 g_A^2 \b_M^{\pq} $
&  $\quad\qquad 0 $ &\\

& $\quad\quad +4g_A g_1 (7\a_M^{\pq} +2\b_M^{\pq}) $
& $\quad\quad -4g_A g_1 (\a_M^{\pq} +2\b_M^{\pq})$
& & \\

& $\quad\quad +g_1^2 (5\a_M^{\pq} -2\b_M^{\pq})]$
& $\quad\quad -g_1^2 (15\a_M^{\pq} +14\b_M^{\pq})] $
&
& \\

& $\quad + m_d [16 g_A^2 \b_M^{\pq}$
& $\quad + m_d [16 g_A^2 \a_M^{\pq} $
& & \\

& $\quad\quad +4 g_A g_1 (\a_M^{\pq} +6\b_M^{\pq})$
& $\quad\quad -4g_A g_1 (3\a_M^{\pq} +2\b_M^{\pq})$
& & \\

& $\quad\quad +g_1^2 (6\b_M^{\pq} -\a_M^{\pq})] \}$
& $\quad\quad -g_1^2  (5\a_M^{\pq} +6\b_M^{\pq})] \}$
&
& \\
\hline

$n$

&  $\qquad\qquad 0 $
&  $\frac{1}{24} \{ m_u [16 g_A^2 \a_M^{\pq}$  
&  $\frac{1}{24} \{ m_u [16 g_A^2 \b_M^{\pq}$
& \\

&
& $\quad\quad - 4g_A g_1 (3\a_M^{\pq} +2\b_M^{\pq})$
& $\quad\quad +4g_A g_1 (\a_M^{\pq} +6\b_M^{\pq})$
& \\

&
& $\quad\quad -g_1^2 (5\a_M^{\pq} +6\b_M^{\pq})] $
& $\quad\quad +g_1^2 (6\b_M^{\pq}-\a_M^{\pq})]$
& \\

& 
& $\quad + m_d [16 g_A^2 \b_M^{\pq}$
& $\quad + m_d [16 g_A^2 \a_M^{\pq}$
& \\

&
& $\quad\quad -4g_A g_1 (\a_M^{\pq} +2\b_M^{\pq})$
& $\quad\quad +4 g_A g_1 (7\a_M^{\pq} +2\b_M^{\pq})$
& \\

& 
& $\quad\quad -g_1^2  (15\a_M^{\pq} +14\b_M^{\pq})] \} $
& $\quad\quad +g_1^2 (5\a_M^{\pq} -2\b_M^{\pq})] \} $
& \\
\hline

\multicolumn{5}{c}{} 
\\
        & $\qquad  \qquad  ju $ & $\qquad \qquad lu $ & $\qquad \qquad jd$ & $\qquad \qquad ld$ \\
\hline
\hline

$p$    

&  $\frac{1}{24} \{ m_u [8 g_A^2 (\a_M^{\pq} +2\b_M^{\pq}) $ 
&  $\frac{1}{24} \{ m_u [8 g_A^2 (\a_M^{\pq} +2\b_M^{\pq}) $  
&  $\frac{1}{12} g_1^2 [ m_u (5\a_M^{\pq} +2\b_M^{\pq}) $ 
&  $\frac{1}{12} g_1^2 [ m_u (5\a_M^{\pq} +2\b_M^{\pq}) $ \\

& $\quad\quad +4 g_A g_1 (\a_M^{\pq} +2\b_M^{\pq}) $
& $\quad\quad +4 g_A g_1 (\a_M^{\pq} +2\b_M^{\pq}) $
& $\quad +m_j (\a_M^{\pq} +4\b_M^{\pq})] $
& $\quad +m_l (\a_M^{\pq} +4\b_M^{\pq})] $ \\

& $\quad\quad +g_1^2 (3\a_M^{\pq} +2\b_M^{\pq})]$
& $\quad\quad +g_1^2 (3\a_M^{\pq} +2\b_M^{\pq})]$
&
& \\

& $\quad +2m_j [8 g_A^2 \a_M^{\pq}$
& $\quad +m_d [8 g_A^2 (\a_M^{\pq} +2\b_M^{\pq})$
& $ $
& $ $ \\

& $\quad\quad +4g_A g_1 \a_M^{\pq} $
& $\quad\quad +4 g_A g_1 (\a_M^{\pq} +2\b_M^{\pq})$
&
&\\

& $\quad\quad +g_1^2 (\a_M^{\pq} +2\b_M^{\pq})]$
& $\quad\quad +g_1^2 (3\a_M^{\pq} +2\b_M^{\pq})]$
&
& \\

& $\quad +m_d [8 g_A^2 (\a_M^{\pq} +2\b_M^{\pq})$
& $\quad +2m_l [8 g_A^2 \a_M^{\pq}$
&
&\\

& $\quad\quad +4 g_A g_1 (\a_M^{\pq} +2\b_M^{\pq})$
& $\quad\quad +4g_A g_1 \a_M^{\pq}] \}$
& $ $
& $ $ \\

& $\quad\quad +g_1^2 (3\a_M^{\pq} +2\b_M^{\pq})] \}$
& $\quad\quad +g_1^2 (\a_M^{\pq} +2\b_M^{\pq})] \}$
&
& \\
\hline

$n$  

&  $\frac{1}{12} g_1^2 [m_d (5\a_M^{\pq} +2\b_M^{\pq}) $ 
&  $\frac{1}{12} g_1^2 [m_d (5\a_M^{\pq} +2\b_M^{\pq}) $ 
&  $\frac{1}{24} \{ m_u [ 8g_A^2 (\a_M^{\pq} +2\b_M^{\pq}) $ 
&  $\frac{1}{24} \{ m_u [ 8g_A^2 (\a_M^{\pq} +2\b_M^{\pq}) $  \\

& $\quad + m_j (\a_M^{\pq} +4\b_M^{\pq})]$
& $\quad + m_l (\a_M^{\pq} +4\b_M^{\pq})] $ 
& $\quad\quad +4g_A g_1 (\a_M^{\pq} +2\b_M^{\pq}) $
& $\quad\quad +4g_A g_1 (\a_M^{\pq} +2\b_M^{\pq}) $\\

&
&
& $\quad\quad +g_1^2 (3\a_M^{\pq} +2\b_M^{\pq})] $
& $\quad\quad +g_1^2 (3\a_M^{\pq} +2\b_M^{\pq})] $ \\

&
&
& $\quad +2m_j [ 8g_A^2 \a_M^{\pq} $ 
& $\quad +m_d [8g_A^2 (\a_M^{\pq} +2\b_M^{\pq}) $  \\

&
&
& $\quad\quad +4g_A g_1 \a_M^{\pq}$
& $\quad\quad +4g_A g_1 (\a_M^{\pq} +2\b_M^{\pq}) $\\

&
&
& $\quad\quad +g_1^2 (\a_M^{\pq} +2\b_M^{\pq})]$
& $\quad\quad +g_1^2 (3\a_M^{\pq} +2\b_M^{\pq})]$ \\

&
&
& $\quad +m_d [8g_A^2 (\a_M^{\pq} +2\b_M^{\pq})$
& $\quad +2m_l [ 8g_A^2 \a_M^{\pq} $\\

&
&
& $\quad\quad +4g_A g_1 (\a_M^{\pq} +2\b_M^{\pq})$
& $\quad\quad +4g_A g_1 \a_M^{\pq}$\\

&
&
& $\quad\quad +g_1^2 (3\a_M^{\pq} +2\b_M^{\pq})] \}$
& $\quad\quad +g_1^2 (\a_M^{\pq} +2\b_M^{\pq})]$\\
\hline

\multicolumn{5}{c}{} 
\\
& \multicolumn{4}{c}{$F^B_{\phi\phi'}$ \phantom{sp}} 
\\
        & $\qquad \qquad \eta_u \eta_u$ & $\qquad \qquad \qquad \eta_u \eta_d $& $\qquad \qquad \eta_d \eta_d $ &\\
\hline
\hline
$p$   
&  $\frac{1}{12} (2g_A +g_1)^2 \times $   
&  $\frac{1}{6} (2 g_A g_1 +g_1^2) \times $  
&  $\frac{1}{12} g_1^2 [ m_u (5\a_M^{\pq} +2\b_M^{\pq}) $ & \\

& $\quad [ m_u (5\a_M^{\pq} +2\b_M^{\pq})$ 
& $\quad [ m_u (5\a_M^{\pq} +2\b_M^{\pq})$
& $\qquad + m_d (\a_M^{\pq} +4\b_M^{\pq})]$
&  \\

& $\quad +m_d (\a_M^{\pq} +4\b_M^{\pq})]$
& $\quad +m_d (\a_M^{\pq} +4\b_M^{\pq})]$
&
& \\

\hline
$n$  
&  $\frac{1}{12} g_1^2[ m_u (\a_M^{\pq} +4\b_M^{\pq})$   
&  $\frac{1}{6} (2 g_A g_1 +g_1^2) \times $  
&  $\frac{1}{12}(2g_A +g_1)^2 \times $ & \\

& $\qquad +m_d (5\a_M^{\pq} +2\b_M^{\pq})]$ 
& $\quad [ m_u (\a_M^{\pq} +4\b_M^{\pq})$
& $\quad [ m_u (\a_M^{\pq} +4\b_M^{\pq})$
& \\

&
& $\quad +m_d(5\a_M^{\pq} +2\b_M^{\pq}) ]$ 
& $\quad +m_d (5\a_M^{\pq} +2\b_M^{\pq})]$
& \\
\end{tabular}
%\end{ruledtabular}
\label{t:NPQQCD-F}
\end{table}
\endgroup
%\end{turnpage}

%\begingroup
%\squeezetable
\begin{table}[ht]
\caption{The coefficients  $G^B_\phi$ and $G^B_{\phi\phi'}$ in \PQCPT. Coefficients are
listed for the nucleons, and for $G^B_\phi$ are grouped into contributions from loop mesons
with mass $m_\phi$, while for $G^B_{\phi\phi'}$ are grouped into contributions from pairs of quark-basis 
$\eta_q$ mesons.}
%\begin{ruledtabular}
\begin{tabular}{l | c c c c }
& \multicolumn{4}{c}{$G^B_\phi \phantom{ap}$} \\
 & $\quad \eta_u \quad$ 
 & $\quad \pi^\pm \quad $ 
 & $\quad \eta_d \quad $ & \\  
\hline
$p$       
 & $\frac{1}{9}(2 m_u + m_d)$ 
 & $\frac{1}{9} ( 13 m_u + 2 m_d)$  
 & $0$ &\\

$n$ 
 & $0$ 
 & $\frac{1}{9} ( 2 m_u + 13 m_d)$ 
 & $\frac{1}{9} (m_u + 2 m_d)$ &\\            

\multicolumn{5}{c}{} 
\\
 & $\quad ju \quad$ 
 & $\quad lu \quad$ 
 & $\quad jd \quad$ 
 & $\quad ld \quad$ \\
\hline
$p$
  & $\frac{1}{9} (m_u + m_d + m_j)$ 
  & $\frac{1}{9}(m_u + m_d + m_l)$  
  & $\frac{2}{9} (2 m_u + m_j)$ 
  & $\frac{2}{9} (2 m_d + m_l)$ \\
$n$
  & $\frac{2}{9}(2 m_d + m_j)$ 
  & $\frac{2}{9}(2 m_d + m_l)$  
  & $\frac{1}{9} (m_u + m_d +  m_j)$ 
  & $\frac{1}{9} (m_u + m_d + m_l)$ \\

\multicolumn{5}{c}{} 
\\
& \multicolumn{4}{c}{$G^B_{\phi\phi'}$ \phantom{sp}} 
\\
  & $\quad \eta_u \eta_u \quad$ 
  & $\quad \eta_u \eta_d \quad $
  & $\quad \eta_d \eta_d \quad$ &\\
\hline
$p$
  & $\frac{2}{9} ( 2 m_u + m_d)$ 
  & $- \frac{4}{9} (2 m_u + m_d)$  
  & $\frac{2}{9} (2 m_u + m_d)$ &\\

$n$
  & $\frac{2}{9}(m_u + 2 m_d)$  
  & $- \frac{4}{9} (m_u + 2 m_d)$  
  & $\frac{2}{9} (m_u + 2 m_d)$ &\\
\end{tabular}
%\end{ruledtabular}
\label{t:NPQQCD-G}
\end{table}

%
%
%  Delta Mass Tables
%
%
%\section{Partially Quenched Delta Coefficients}\label{ap:D}

\begin{table}[ht]
\caption{The coefficients $A^T_\phi$, $A^T_{\phi\phi'}$, $B^T_\phi$
  and $B^T_{\phi\phi'}$ in \PQCPT. Coefficients are listed for the
  delta states $T$, and for $A^T_\phi$ and $B^T_\phi$ are grouped into
  contributions from loop mesons with mass $m_\phi$, while for
  $A^T_{\phi\phi'}$ and $B^T_{\phi\phi'}$ are grouped into
  contributions from pairs of quark-basis $\eta_q$ mesons.}
%\begin{ruledtabular}
\begin{tabular}{l | c c c c c c c | c c c }
 & \multicolumn{7}{c|}{$A^T_\phi \phantom{ap}$} & \multicolumn{3}{c}{$A^T_{\phi\phi'}$ \phantom{sp}} \\
    & $\quad \eta_u \quad$ & $\quad \pi^\pm \quad $ & $\quad  \eta_d \quad $ 
    & $ \quad ju \quad$ & $ \quad lu \quad$ 
    & $\quad jd \quad$  & $\quad ld \quad$ 
    & $\quad \eta_u \eta_u \quad $ & $\quad \eta_u \eta_d\quad $   & $\quad \eta_d \eta_d\quad$ \\
\hline
$\D^{++}$  &  $\frac{2}{3}$ & $0$  & $0$  
           &  $\frac{1}{3}$ & $\frac{1}{3}$  
           &  $0$ & $0$
           &  $1$ & $0$ & $0$ \\

$\D^+$     &  $\frac{2}{9}$ & $\frac{4}{9}$  & $0$  
           &  $\frac{2}{9}$ & $\frac{2}{9}$  
           &  $\frac{1}{9}$ & $\frac{1}{9}$
           &  $\frac{4}{9}$ & $\frac{4}{9}$ & $\frac{1}{9}$ \\

$\D^0$     &  $0$ & $\frac{4}{9}$  & $\frac{2}{9}$  
           &  $\frac{1}{9}$ & $\frac{1}{9}$  
           &  $\frac{2}{9}$ & $\frac{2}{9}$
           &  $\frac{1}{9}$ & $\frac{4}{9}$ & $\frac{4}{9}$ \\

$\D^-$     &  $0$ & $0$  & $\frac{2}{3}$  
           &  $0$ & $0$  
           &  $\frac{1}{3}$ & $\frac{1}{3}$
           &  $0$ & $0$ & $1$ \\
 
\end{tabular}

\bigskip

\begin{tabular}{l | c c c c c c c | c c c }
 & \multicolumn{7}{c|}{$B^T_\phi \phantom{ap}$} & \multicolumn{3}{c}{$B^T_{\phi\phi'}$ \phantom{sp}} \\
    & $\quad \eta_u \quad$ & $\quad \pi^\pm \quad $ & $\quad \eta_d \quad $ 
    & $ \quad ju \quad$ & $ \quad lu \quad$ 
    & $\quad jd \quad$  & $\quad ld \quad$ 
    & $\quad \eta_u \eta_u \quad $ & $\quad \eta_u \eta_d \quad $   & $\quad \eta_d \eta_d \quad$ \\
\hline
$\D^{++}$       &  $-\frac{2}{3}$ & $0$  & $0$  
           &  $\frac{2}{3}$ & $\frac{2}{3}$  
           &  $0$ & $0$
           &  $0$ & $0$ & $0$ \\

$\D^+$     &  $-\frac{2}{9}$ & $-\frac{4}{9}$  & $0$  
           &  $\frac{4}{9}$ & $\frac{4}{9}$  
           &  $\frac{2}{9}$ & $\frac{2}{9}$
           &  $\frac{2}{9}$ & $-\frac{4}{9}$ & $\frac{2}{9}$ \\

$\D^0$     &  $0$ & $-\frac{4}{9}$  & $-\frac{2}{9}$  
           &  $\frac{2}{9}$ & $\frac{2}{9}$  
           &  $\frac{4}{9}$ & $\frac{4}{9}$
           &  $\frac{2}{9}$ & $-\frac{4}{9}$ & $\frac{2}{9}$ \\

$\D^-$     &  $0$ & $0$  & $-\frac{2}{3}$  
           &  $0$ & $0$  
           &  $\frac{2}{3}$ & $\frac{2}{3}$
           &  $0$ & $0$ & $0$ \\
 
\end{tabular}
%\end{ruledtabular}
\label{t:PQQCD-A}
\end{table}
%\endgroup

%\begingroup
%\squeezetable
\begin{table}[ht]
\caption{The coefficients $C^T_\phi$, $C^T_{\phi\phi'}$, $D^T_\phi$
  and $D^T_{\phi\phi'}$ in \PQCPT\ are listed for the delta states
  $T$, and for $C^T_\phi$ and $D^T_\phi$ are grouped into
  contributions from loop mesons with mass $m_\phi$, while for
  $C^T_{\phi\phi'}$ and $D^T_{\phi\phi'}$ are grouped into
  contributions from pairs of quark-basis $\eta_q$ mesons. If a
  particular meson or pair of flavor-neutral mesons is not listed,
  then the value of the coefficient is zero for all deltas. The
  coefficients $\ol C_\phi$, $\ol C_{\phi \phi'}$, $\ol D_\phi$ and
  $\ol D_{\phi \phi'}$  are identical for all nucleons and deltas and
  appear in Table \ref{t:NPQQCD-C}.}

%\begin{ruledtabular}
\begin{tabular}{l | c c c c | c c}
& \multicolumn{4}{c|}{$C^T_\phi $} & \multicolumn{2}{c}{$C^T_{\phi\phi'}$} \\
      & $\quad ju \quad$ & $\quad lu \quad $ & $\quad jd \quad $  & $\quad ld \quad$ & $\quad \eta_u \eta_u \quad$ & $\quad \eta_d \eta_d \quad$ \\
\hline
$\D^{++}$       
           &  $m_u + m_j$ & $m_u + m_l$  & $0$ & $0$ 
           &  $2 m_u$ & $0$ \\

$\D^+$ 
           &  $\frac{2}{3}( m_u + m_j)$ & $\frac{2}{3}(m_u + m_l)$  & $\frac{1}{3}(m_d +  m_j)$ & $\frac{1}{3}(m_d + m_l)$
           &  $\frac{4}{3} m_u$ & $\frac{2}{3} m_d$ \\

$\D^0$    
           &  $\frac{1}{3}(m_u + m_j)$ & $\frac{1}{3} (m_u + m_l)$  & $\frac{2}{3}( m_d + m_j)$ & $\frac{2}{3} (m_d + m_l)$ 
           &  $\frac{2}{3} m_u$ & $\frac{4}{3} m_d$ \\

$\D^-$ 
           &  $0$ & $0$  & $m_d + m_j$ & $m_d + m_l$ 
           &  $0$ & $2 m_d$ 
\end{tabular}

\bigskip

\begin{tabular}{l | c c c c | c c}
  & \multicolumn{4}{c|}{$D^T_\phi \phantom{ap}$} 
  & \multicolumn{2}{c}{$D^T_{\phi\phi'}$ \phantom{sp}} \\
  & $\quad ju \quad$ 
  & $\quad lu \quad $ 
  & $\quad jd \quad $  
  & $\quad ld \quad$ 
  & $\quad \eta_u \eta_u \quad$ 
  & $\quad \eta_d \eta_d \quad$ \\
\hline
$\D^{++}$       
  & $1$ 
  & $1$  
  & $0$ 
  & $0$ 
  & $1$ 
  & $0$ \\

$\D^+$ 
  & $\frac{2}{3}$ 
  & $\frac{2}{3}$  
  & $\frac{1}{3}$ 
  & $\frac{1}{3}$
  & $\frac{2}{3}$ 
  & $\frac{1}{3}$ \\

$\D^0$    
  & $\frac{1}{3}$ 
  & $\frac{1}{3}$  
  & $\frac{2}{3}$ 
  & $\frac{2}{3}$ 
  & $\frac{1}{3}$ 
  & $\frac{2}{3}$ \\

$\D^-$ 
  & $0$ 
  & $0$  
  & $1$ 
  & $1$ 
  & $0$ 
  & $1$ \\
\end{tabular}
%\end{ruledtabular}
\label{t:PQQCD-C}
\end{table}
%\endgroup

\begin{table}[ht]
\caption{The coefficients  $F^T_\phi$ and $F^T_{\phi\phi'}$ in \PQCPT. Coefficients are
listed for the delta states $T$, and for $F^T_\phi$ are grouped into contributions from loop mesons
with mass $m_\phi$, while for $F^T_{\phi\phi'}$ are grouped into contributions from pairs of quark-basis 
$\eta_q$ mesons.}
%\begin{ruledtabular}
\begin{tabular}{l | c c c c }
& \multicolumn{4}{c}{$F^T_\phi \phantom{ap}$} \\
      & $\quad \eta_u \quad$ & $\quad \pi^\pm \quad $ & $\quad \eta_d \quad $ & \\  
\hline
$\D^{++}$       
           &  $2 m_u$ & $0$  & $0$ &\\

$\D^+$ 
           &  $\frac{2}{9}(2 m_u + m_d)$ & $\frac{4}{9} ( 2 m_u + m_d)$  & $0$ &\\            

$\D^0$    
           &  $0$ & $\frac{4}{9} ( m_u + 2 m_d )$  & $\frac{2}{9}(m_u + 2 m_d)$ &\\ 

$\D^-$ 
           &  $0$ & $0$  & $2 m_d$ & \\
\multicolumn{5}{c}{} 
\\
        & $\quad ju \quad$ & $\quad lu \quad$ & $\quad jd \quad$ & $\quad ld \quad$ \\
\hline
$\D^{++}$    &  $\frac{1}{3} (2 m_u + m_j)$ & $\frac{1}{3}(2 m_u + m_l)$  & $0$ & $0$ \\
$\D^+$       &  $\frac{2}{9}(m_u + m_d + m_j)$ & $\frac{2}{9}(m_u + m_d + m_l)$  & $\frac{1}{9} (2 m_u + m_j)$ & $\frac{1}{9} (2 m_u + m_l)$ \\
$\D^0$       &  $\frac{1}{9}(2 m_d + m_j)$ & $\frac{1}{9}(2 m_d + m_l)$  & $\frac{2}{9}(m_u + m_d + m_j)$ & $\frac{2}{9}(m_u + m_d + m_l)$ \\
$\D^-$       &  $0$ & $0$  & $\frac{1}{3}(2 m_d + m_j)$ & $\frac{1}{3} (2 m_d + m_l)$ \\
\multicolumn{5}{c}{} 
\\
& \multicolumn{4}{c}{$F^T_{\phi\phi'}$ \phantom{sp}} 
\\
        & $\quad \eta_u \eta_u \quad$ & $\quad \eta_u \eta_d \quad $& $\quad \eta_d \eta_d \quad$ &\\
\hline
$\D^{++}$    &  $3 m_u$ & $0$  & $0$ &\\
$\D^+$       &  $\frac{4}{9}(2 m_u + m_d)$   & $\frac{4}{9}(2 m_u + m_d)$  & $\frac{1}{9}(2 m_u + m_d)$ & \\
$\D^0$       &  $\frac{1}{9} ( m_u + 2 m_d )$  & $\frac{4}{9}(m_u + 2 m_d)$  & $\frac{4}{9}(m_u + 2 m_d)$ & \\
$\D^-$       &  $0$ & $0$  & $3 m_d$ & 
\end{tabular}
%\end{ruledtabular}
\label{t:PQQCD-F}
\end{table}
%\endgroup

\begingroup
\squeezetable

\begin{table}[ht]
\caption{The coefficients $E^T_\phi$ and $E^T_{\phi\phi'}$ in \PQCPT. Coefficients are
listed for the delta states $T$, and for $E^T_\phi$ are grouped into contributions from loop mesons
with mass $m_\phi$, while for $E^T_{\phi\phi'}$ are grouped into contributions from pairs of quark-basis 
$\eta_q$ mesons.}
%\begin{ruledtabular}
\begin{tabular}{l | c c c | c  c c}
& \multicolumn{3}{c|}{$E^T_\phi \phantom{ap}$} & \multicolumn{3}{c}{$E^T_{\phi\phi'}$ \phantom{sp}} \\
      & $\quad \eta_u \quad$ & $\quad \pi^\pm \quad $ & $\quad \eta_d \quad $  & $\quad \eta_u \eta_u \quad$ & $\quad \eta_u \eta_d \quad $& $\quad \eta_d \eta_d \quad$ \\
\hline
$\D^{++}$       
           &  $1$ & $0$  & $0$ 
           & $1$  & $0$  & $0$ \\

$\D^+$ 
           &  $\frac{1}{3}$ & $\frac{2}{3}$  & $0$ 
           &  $\frac{1}{3}$ & $\frac{2}{3}$  & $0$ \\

$\D^0$    
           &  $0$ & $\frac{2}{3}$  & $\frac{1}{3}$ 
           &  $0$ & $\frac{2}{3}$  & $\frac{1}{3}$ \\

$\D^-$ 
           &  $0$ & $0$  & $1$ 
           &  $0$ & $0$ & $1$ 
\end{tabular}
%\end{ruledtabular}
\label{t:PQQCD-E}
\end{table}

\begin{table}[ht]
\caption{The coefficients $G^T_\phi$ and $G^T_{\phi\phi'}$ in \PQCPT. Coefficients are
listed for the delta states $T$, and for $G^T_\phi$ are grouped into contributions from loop mesons
with mass $m_\phi$, while for $G^T_{\phi\phi'}$ are grouped into contributions from pairs of quark-basis 
$\eta_q$ mesons.}
%\begin{ruledtabular}
\begin{tabular}{l | c c c c }
& \multicolumn{4}{c}{$G^T_\phi \phantom{ap}$} \\
      & $\quad \eta_u \quad$ & $\quad \pi^\pm \quad $ & $\quad \eta_d \quad $ & \\  
\hline
\hline

$\D^{++}$       
&  $- \frac{4}{3} m_u (\a_M^{(PQ)} + \b_M^{(PQ)})$ 
&  $0$  
&  $0$ &\\

\hline

$\D^+$
&  $-\frac{4}{27}[m_u (\a_M^{(PQ)} + 4 \b_M^{(PQ)})$ 
&  $-\frac{4}{27}[m_u ( 5 \a_M^{(PQ)} + 2 \b_M^{(PQ)})$  
&  $0$ &\\            

& $\qquad + m_d (2 \a_M^{(PQ)} - \b_M^{(PQ)})]$
& $\qquad + m_d (\a_M^{(PQ)} + 4\b_M^{(PQ)})]$
& $$ &\\

\hline
$\D^0$    
&  $0$ 
&  $-\frac{4}{27}[m_u(\a_M^{(PQ)} + 4\b_M^{(PQ)})$  
&  $-\frac{4}{27}[m_u(2 \a_M^{(PQ)} - \b_M^{(PQ)})$ &\\

& $$
& $\qquad + m_d (5 \a_M^{(PQ)} + 2 \b_M^{(PQ)})]$
& $\qquad + m_d (\a_M^{(PQ)} + 4 \b_M^{(PQ)})]$ &\\
 
\hline
$\D^-$
&  $0$ 
&  $0$  
&  $- \frac{4}{3} m_d (\a_M^{(PQ)} + \b_M^{(PQ)})$ & \\

\multicolumn{5}{c}{} 
\\
        & $\quad ju \quad$ & $\quad lu \quad$ & $\quad jd \quad$ & $\quad ld \quad$ \\
\hline
\hline

$\D^{++}$    
&  $\frac{2}{9}[m_u (5 \a_M^{(PQ)} + 2 \b_M^{(PQ)}) $ 
&  $\frac{2}{9}[m_u (5 \a_M^{(PQ)} + 2 \b_M^{(PQ)})$  
&  $0$ 
&  $0$ \\

& $\qquad + m_j ( \a_M^{(PQ)} + 4 \b_M^{(PQ)})]$
& $\qquad + m_l ( \a_M^{(PQ)} + 4 \b_M^{(PQ)})]$
& $$
& $$ \\

\hline

$\D^+$  
&  $\frac{2}{27} [ m_u (5 \a_M^{(PQ)} + 2 \b_M^{(PQ)})$ 
&  $\frac{2}{27} [ m_u (5 \a_M^{(PQ)} + 2 \b_M^{(PQ)})$  
&  $\frac{2}{27} [ m_u (5 \a_M^{(PQ)} + 2 \b_M^{(PQ)})$ 
&  $\frac{2}{27} [ m_u (5 \a_M^{(PQ)} + 2 \b_M^{(PQ)})$ \\

& $\qquad + m_d (5 \a_M^{(PQ)} + 2 \b_M^{(PQ)})$
& $\qquad + m_d (5 \a_M^{(PQ)} + 2 \b_M^{(PQ)})$
& $\qquad + m_j ( \a_M^{(PQ)} + 4 \b_M^{(PQ)})]$
& $\qquad + m_l ( \a_M^{(PQ)} + 4 \b_M^{(PQ)})]$ \\

& $\qquad + 2 m_j ( \a_M^{(PQ)} + 4 \b_M^{(PQ)})]$
& $\qquad + 2 m_l ( \a_M^{(PQ)} + 4 \b_M^{(PQ)})]$
& $$
& $$ \\

\hline

$\D^0$    
&  $\frac{2}{27} [ m_d (5 \a_M^{(PQ)} + 2 \b_M^{(PQ)})$ 
&  $\frac{2}{27} [ m_d (5 \a_M^{(PQ)} + 2 \b_M^{(PQ)})$  
&  $\frac{2}{27} [ m_u (5 \a_M^{(PQ)} + 2 \b_M^{(PQ)})$ 
&  $\frac{2}{27} [ m_u (5 \a_M^{(PQ)} + 2 \b_M^{(PQ)})$ \\

& $\qquad + m_j ( \a_M^{(PQ)} + 4 \b_M^{(PQ)})]$
& $\qquad + m_l ( \a_M^{(PQ)} + 4 \b_M^{(PQ)})]$
& $\qquad + m_d (5 \a_M^{(PQ)} +  2 \b_M^{(PQ)})$
& $\qquad + m_d (5 \a_M^{(PQ)} + 2 \b_M^{(PQ)})$ \\

& $$
& $$
& $\qquad + 2 m_j ( \a_M^{(PQ)} + 4 \b_M^{(PQ)})]$
& $\qquad + 2 m_l ( \a_M^{(PQ)} + 4 \b_M^{(PQ)})]$ \\

\hline

$\D^-$   
&  $0$ 
&  $0$  
&  $\frac{2}{9}[m_d (5 \a_M^{(PQ)} + 2 \b_M^{(PQ)})$ 
&  $\frac{2}{9}[m_d (5 \a_M^{(PQ)} + 2 \b_M^{(PQ)})$ \\

& $$
& $$
& $\qquad + m_j ( \a_M^{(PQ)} + 4 \b_M^{(PQ)})]$
& $\qquad + m_l ( \a_M^{(PQ)} + 4 \b_M^{(PQ)})]$ \\
\multicolumn{5}{c}{} 
\\
& \multicolumn{4}{c}{$G^T_{\phi\phi'}$ \phantom{sp}} 
\\
        & $\quad \eta_u \eta_u \quad$ & $\quad \eta_u \eta_d \quad $& $\quad \eta_d \eta_d \quad$ &\\
\hline
\hline
$\D^{++}$   
&  $0$ 
&  $0$  
&  $0$ &\\

\hline
$\D^+$  
&  $\frac{2}{27}[m_u (5 \a_M^{(PQ)} + 2 \b_M^{(PQ)})$   
&  $-\frac{4}{27}[ m_u (5 \a_M^{(PQ)} + 2 \b_M^{(PQ)})$  
&  $\frac{2}{27} [m_u(5 \a_M^{(PQ)} + 2 \b_M^{(PQ)})$ & \\

& $\qquad + m_d (\a_M^{(PQ)} + 4 \b_M^{(PQ)})]$ 
& $\qquad + m_d (\a_M^{(PQ)} + 4 \b_M^{(PQ)})]$
& $\qquad + m_d (\a_M^{(PQ)} + 4 \b_M^{(PQ)})]$ & \\
\hline

$\D^0$ 
&  $\frac{2}{27}[m_u (\a_M^{(PQ)} + 4 \b_M^{(PQ)})$  
&  $-\frac{4}{27}[ m_u (\a_M^{(PQ)} + 4 \b_M^{(PQ)})$  
&  $\frac{2}{27}[m_u (\a_M^{(PQ)} + 4 \b_M^{(PQ)})$ & \\

& $\qquad + m_d (5 \a_M^{(PQ)} + 2 \b_M^{(PQ)})]$
& $\qquad + m_d (5 \a_M^{(PQ)} + 2 \b_M^{(PQ)})]$
& $\qquad + m_d (5 \a_M^{(PQ)} + 2 \b_M^{(PQ)})]$ & \\

\hline

$\D^-$  
&  $0$ 
&  $0$  
&  $0$ & 
\end{tabular}
%\end{ruledtabular}
\label{t:PQQCD-G}
\end{table}
\endgroup

\end{document}